\documentclass[12pt,letterpaper]{article}
\usepackage[footskip=0.75in]{geometry}

\usepackage{amsmath,amssymb,chemarr}
\usepackage{changepage}

\usepackage{textcomp,marvosym}

\usepackage{cite}

\usepackage{nameref,hyperref}

\usepackage{todonotes}
\hypersetup{
    colorlinks=true,
    linkcolor=blue,
    citecolor=blue,
    urlcolor=blue
}
\graphicspath{{figs/}}

\usepackage[right]{lineno}

\usepackage{microtype}
\DisableLigatures[f]{encoding = *, family = * }

\usepackage{array}

\newcolumntype{+}{!{\vrule width 2pt}}

\newlength\savedwidth

\raggedright
\usepackage[aboveskip=1pt,labelfont=bf,labelsep=period,justification=raggedright,singlelinecheck=off]{caption}

\newcommand{\differential}[1]{{\mathrm d } #1} 

\makeatletter
\renewcommand{\@biblabel}[1]{\quad#1.}
\makeatother

\usepackage{pifont}
\newcommand{\xmark}{\ding{55}}%

\usepackage{algorithm,algorithmic}

\usepackage{lastpage,fancyhdr,graphicx}
\usepackage{epstopdf}
\fancyheadoffset[L]{2.25in}
\fancyfootoffset[L]{2.25in}
\begin{document}
\vspace*{0.2in}

\begin{flushleft}
{\Large
\textbf\newline{Inference for stochastic reaction networks via logistic regression} 
}
\newline
\\
Boseung
Choi\textsuperscript{1\dag},
Hye-Won
Kang\textsuperscript{2\dag},
Grzegorz
A. Rempala\textsuperscript{3\dag*},
\\
\bigskip
\textbf{1} Korea University Sejong Campus, Sejong, South Korea
\\
\textbf{2} University of Maryland, Baltimore County, Baltimore MD, USA
\\
\textbf{3} The Ohio State University, Columbus, OH, USA
\\
\bigskip

%
%

\dag These authors contributed equally to this work.




* rempala.3@osu.edu

\end{flushleft}
\section*{Abstract}
Identifying network structure and inferring parameters are central challenges in modeling chemical reaction networks. In this study, we propose likelihood-based methods grounded in multinomial logistic regression to infer both stoichiometries and network connectivity structure from full time-series trajectories of stochastic chemical reaction networks. When complete molecular count trajectories are observed for all species, stoichiometric coefficients are identifiable, provided each reaction occurs at least once during the observation window. However, identifying catalytic species remains difficult,  as their molecular counts remain unchanged before and after each reaction event. Through three illustrative stochastic models involving catalytic interactions in open networks, we demonstrate that the logistic regression framework, when applied properly, can recover the full network structure, including stoichiometric relationships. We further apply Bayesian logistic regression to estimate model parameters in real-world epidemic settings, using the COVID-19 outbreak in the Greater Seoul area of South Korea as a case study. Our analysis focuses on a Susceptible–Infected–Recovered (SIR) network model that incorporates demographic effects. To address the challenge of partial observability, particularly the availability of data only for the infectious subset of the population, we develop a method that integrates Bayesian logistic regression with differential equation models. This approach enables robust inference of key SIR parameters from observed COVID-19 case trajectories. Overall, our findings demonstrate that simple, likelihood-based techniques such as logistic regression can recover meaningful mechanistic insights from both synthetic and empirical time-series data.



\section*{Introduction}
Identifying network structures from time-series trajectories is a fundamental challenge across scientific disciplines, including systems biology, epidemiology, and chemical kinetics. A variety of methods have been proposed to address this problem, depending on factors such as the completeness of the observations, the temporal resolution of the data (continuous vs. discrete), and the presence of noise.

In deterministic mass action kinetics, different reaction networks may produce identical dynamics, and model parameters may not be uniquely identifiable~\cite{Craciun:2008:ICR}. 
Ordinary differential equations (ODEs) with polynomial right-hand sides are commonly used to select relevant terms and estimate parameters from time-series data, employing regression~\cite{Karnaukhov:2007:NMM,Burnham:2008:ICR,Brunton:2016:DGE} and machine learning techniques~\cite{Brunton:2016:DGE}. These methods have been extended to infer reaction networks under mass action kinetics~\cite{Hoffmann:2019:RSD}, and further adopted for fully or partially observable species using Bayesian approaches~\cite{Jiang:2022:IDM}.

A similar approach was developed in the stochastic setting, assuming all species and reactions' occurrences are fully observable.  Polynomial propensity functions up to a specified degree were treated as base functions, and parameters within these functions were inferred by maximizing the likelihood function~\cite{Zhang:2019:LCR}.
It has also been shown that network structure and parameters can be uniquely identified when transition rates over a sufficiently large state space or enough full-time trajectories are available ~\cite{Enciso:2021:ISM}.

When the network structure is known, parameter estimation still requires time-series data. 
With full observability, Bayesian inference using Markov Chain Monte Carlo (MCMC) methods with reversible jump algorithms or block updating is effective~\cite{Boys:2008:BID}. For partially observed systems, these methods can be adapted~\cite{Boys:2008:BID},
or uniformization techniques can approximate conditional distributions~\cite{Choi:2011:IDO}. MCMC method using distributed time delay can be also applied to partially observed systems to infer the parameters of the system as well as the parameters of the time delay~\cite{Hong:2023:DDE}. 

In this paper, we focus on simple likelihood-based approaches for network inference and parameter estimation in settings where full trajectory data is available. We consider stochastic reaction networks in a well-mixed environment, where species undergo creation, decay, or interaction over time. When trajectories are fully observed, they capture the temporal dynamics of net changes in species counts. However, since different reactions can produce identical net changes, recovering the true network structure remains challenging. Even after identifying the correct structure, estimating the rates of individual reactions poses additional complexity.

To address these challenges, we organize the network inference task into two sequential stages:

\begin{enumerate}
    \item \textbf{Network structure identification}: The first stage focuses on reconstructing the underlying reaction network from observed data by determining which species abundances influence the occurrence of specific reactions. We show that logistic regression can be effectively applied to fully observed trajectory data, allowing us to identify the species whose elevated abundance is predictive of particular reaction events.

   \item \textbf{Parameter estimation}: After establishing the network structure, the second stage aims to estimate the \textit{kinetic reaction rate constants}, which characterize the magnitude and frequency of interactions within the system. For this purpose, we again employ a logistic regression framework, enhanced with an offset correction, to infer the reaction rate parameters from the same trajectory data.
\end{enumerate}

Building on this two-step approach, we structure the paper as follows. In the Methods section, we introduce the formalism of chemical reaction networks and present a general framework that uses logistic regression for network inference and Bayesian inference for the logistic regression model.  

In the Results section, we first demonstrate the structure inference step using three numerical examples: the Togashi–Kaneko (TK) model—an autocatalytic chemical reaction network \cite{Togashi:2001:TID}; a model of the heat shock response \cite{Linder:2020:SLM}; and a stochastic SIR model with demography—a widely used framework in epidemiology \cite{Andersson2000}. 

We then turn to parameter estimation using real-world data. Specifically, we apply our methods to daily new infection count data from the early stages of the COVID-19 epidemic in the Greater Seoul area. Using a Susceptible–Infected–Recovered (SIR) model with demographic dynamics, we demonstrate how our logistic  framework offers a practical approach for estimating key SIR model epidemiological parameters, including transmission and recovery rates.

In the Discussion section,  we conclude by summarizing our results, evaluating the effectiveness of the proposed methodology, and outlining directions for future research. 

By decomposing the problem into two distinct steps—network structure inference followed by parameter estimation—we propose a practical and interpretable approach for reconstructing reaction networks from time-series data. Our findings show that even simple likelihood-based techniques, when applied appropriately, can reliably uncover underlying network structures and produce plausible parameter estimates in both synthetic and real-world settings.

\section*{Methods}

\subsection*{Network structure identification}

We consider a chemical reaction network (CRN) involving $m$ chemical reactions and $s$ chemical species. 
Denote the $i$-th species as $A_i$, for $i=1,2,\cdots, s$. The stoichiometric coefficient $\nu_{ik}$ ($\nu_{ik}'$), for $i=1,2,\cdots, s$, $k=1,2,\cdots,m$, represents the number of molecules of species $A_i$ that is consumed (or produced) in the $k$-th reaction. Then, the CRN is given as
\begin{equation}
\label{general_crn}
\sum_{i=1}^s \nu_{ik} A_i \longrightarrow \sum_{i=1}^s \nu_{ik}'A_i,
\quad \mbox{for $k=1,2,\cdots, m$.}
\end{equation} The species on the left-hand side of a reaction are often referred to as \emph{reactants}, while those on the right-hand side are referred to as \emph{products}.

We construct a continuous-time Markov jump process to model the stochastic chemical reaction network (CRN) described in \eqref{general_crn}. For a more detailed introduction to the Markov chain model of the CRN, see~\cite{Anderson:2015:SAB}. Full trajectory data for the stochastic model are then generated through repeated simulation using Gillespie's Stochastic Simulation Algorithm~\cite{Gillespie:1977:ESS,Gillespie:2007:SSC}. These synthetic trajectory datasets are used for inferring the network structure, assuming that all species are observable at every reaction event. Note that a large dataset is not required--accurate inference of the network structure is achievable with a modest number of simulations (e.g., 10 trajectories), as demonstrated in the Results section.

Since reaction events are observed at all time points, the stoichiometry of each reaction can be identified from the dataset, provided that all reactions occur at least once during the observation period. Once the stoichiometry is known, the net changes in molecular counts can be determined; however, the presence of catalysts in the reactions cannot be inferred from this information alone, as catalysts do not alter stoichiometry.

Since reaction events are observed at all time points, the stoichiometry of each reaction can be identified from the dataset, provided that all reactions occur at least once during the observation period. Once the stoichiometry is known, the net changes in molecular counts can be determined; however, the presence of catalysts in the reactions cannot be inferred from this information alone, as catalysts do not alter stoichiometry.

To address this problem, we employ a multinomial logistic regression approach. Let $ X_i $ denote the molecular count of the $ i $-th species. We assume that the log ratio of the probability of occurrence of the $ k $-th reaction relative to a reference reaction is a linear function of the molecular counts of the species. The multinomial logistic regression model~\cite{Hosmer:2013:ALR} is then given by
\begin{equation} \label{multinomial_logistic_model}
    \log\left(\frac{P(Y = k)}{P(Y = r)}\right) = \alpha_k + \sum_{i=1}^s \beta_{ki} X_{i},
    \qquad \mbox{for $k=1,2,\cdots,m,\,k\neq r$.}
\end{equation}
Here $Y$ 
denotes a categorical random variable indicating the reaction type, with the 
$r$-th reaction chosen as the reference category in the multinomial logistic regression model. Each stoichiometric type is treated as a nominal category. For each observed reaction event, we pair the molecular counts of all species immediately preceding the event with the corresponding stoichiometric category. Using this combined dataset, we fit a multinomial logistic regression model, typically selecting a production reaction as the reference category.

The estimated regression coefficients capture how species abundances affect the likelihood of each reaction type. The \emph{sign} of each coefficient aids interpretation: a positive value indicates that higher species counts increase the reaction likelihood, suggesting a reactant role. Species with no net change, such as catalysts, can still be identified through their positive association with reaction propensity despite unchanged molecular counts. This framework thus enables systematic detection of reactant species, including catalytic participants.

\subsection*{Parameter estimation} 

For parameter estimation, we adopt a slightly more general—and thus more flexible—form of the model in \eqref{multinomial_logistic_model} and employ a Bayesian framework, wherein the regression coefficients and any additional unknown terms are jointly inferred through  their posterior distributions.

We consider a CRN as described in \eqref{general_crn}, and assume that a subset of the chemical species is observable—that is, full trajectory data are available for these observable species, while the remaining species are unobserved. For the reactions involving unobserved species, we assume that approximate trajectory information may be available from other sources, such as some auxiliary dynamical models. 

We further generalize the model in \eqref{multinomial_logistic_model} by assuming that the log-ratio of the probability of the $k$-th reaction, relative to a reference reaction, is a linear function of the molecular counts of the observed species, augmented by an offset term. The resulting multinomial logistic regression model takes the form
\begin{equation} \label{multinomial_logistic_model_offset}
    \log\left(\frac{P(Y_j = k)}{P(Y_j = r)}\right) = \alpha_{k} + \sum_{i\in \mathcal{O}} \beta_{ki} X_{ij} + \mathrm{offset\,}_{kj},
\end{equation}
where 
$Y_j$ denotes a categorical random variable indicating the reaction type observed at the 
$j$-th observation time, corresponding to the 
$j$-th jump of the process.
The variable $X_{ij}$ denotes the molecular count of the $i$-th species at the $j$-th observation time.
The offset term $\rm{offset\,}_{kj}$ captures the influence of unobserved species on the occurrence of the $k$-th reaction and can be approximated from the behavior of a related dynamical system at the $j$-th time point. 
The probability $P(Y_j=k)$ represents the probability that the $k$-th reaction occurs at the $j$-th observation time, i.e., that the observed reaction at the $j$-th time point is the $k$-th reaction.

Let $\theta$ denote the vector of unknown parameters to be estimated. The likelihood function for the observed data $Y$ can be expressed as
\begin{eqnarray}\label{eq:lkhd}
L(\theta \mid Y) = \prod_{j} \prod_{k=1}^m P(Y_j = k)^{\mathbb{I}_{\{y_j = k\}}},
\end{eqnarray}
where $\mathbb{I}_{\{y_j = k\}}$ is the indicator function and $y_j \in \{1,2,\dots,m\}$ denotes the observed outcome of the random variable $Y_j$ at time $j$.  
The posterior distribution of the parameters, $\pi(\theta \mid Y)$, is then given up to a normalizing constant by
\begin{eqnarray}\label{eq:propto}
\pi(\theta \mid Y) \propto L(\theta \mid Y)\,\pi(\theta),
\end{eqnarray}
where $\pi(\theta)$ represents the prior distribution encoding information about $\theta$ before observing the data.

To carry out Bayesian estimation, we employ Markov Chain Monte Carlo (MCMC) methods to approximate samples from the posterior distribution \( \pi(\theta \mid Y) \).  
Specifically, we use the robust adaptive Metropolis (RAM) algorithm~\cite{Vihola:2012:RAM}, which dynamically adjusts the proposal covariance matrix to achieve an optimal acceptance rate, improving sampling efficiency and reducing the need for manual tuning (see also~\cite{Andrieu:2008:AMCMC} for a broader discussion of adaptive MCMC).  
The resulting samples enable computation of posterior summaries and facilitate probabilistic inference for the model parameters. For additional discussion and illustrative examples, see~\cite{Wilkinson:2018:SMS}.

\section*{Results}


\subsection*{Network structure identification}

In the following, we illustrate our network identification approach using three synthetic yet biologically motivated examples of reaction networks: the well-studied Togashi–Kaneko (TK) model, a canonical heat shock response model—both of which have been previously analyzed in the literature (see, e.g., \cite{Togashi:2001:TID} and \cite{Linder:2020:SLM})—and the SIR model with demography \cite{Andersson2000}, which is considered a foundational workhorse in epidemic modeling.

\subsubsection*{The Togashi-Kaneko (TK) model}
The TK model consists of a cycle of autocatalytic reactions, along with inflow and outflow reactions.
It was first introduced by Togashi and Kaneko~\cite{Togashi:2001:TID}. More recently, the long time behavior of the TK model, as well as its generalizations, has been studied~\cite{Bibbona:2020:SDS}.
The reaction network for $m$ species $\{A_i\}_{i=1}^{m}$ is given by:

\begin{align}
    A_i + A_{i+1} &\xrightarrow{\kappa_i} 2A_{i+1}, \quad i = 1,2,\dots,m, \quad \text{where } A_{m+1} = A_1. \\
    \emptyset &\xrightleftharpoons[\delta_i]{\lambda_i} A_i, \quad i = 1, \dots, m.
\end{align}

This model plays a significant role in molecular-level stochastic kinetics, as it exhibits discrete-induced transitions that are absent in deterministic formulations. Consequently, the TK model may serve as an illustrative test case for network inference using a logistic regression approach.

We specifically consider the case with two species ($ m = 2 $) to evaluate the method's ability to identify all reacting species (reactants) whose molecular counts influence the reaction rates.

\begin{figure}
\includegraphics[width=1\textwidth]{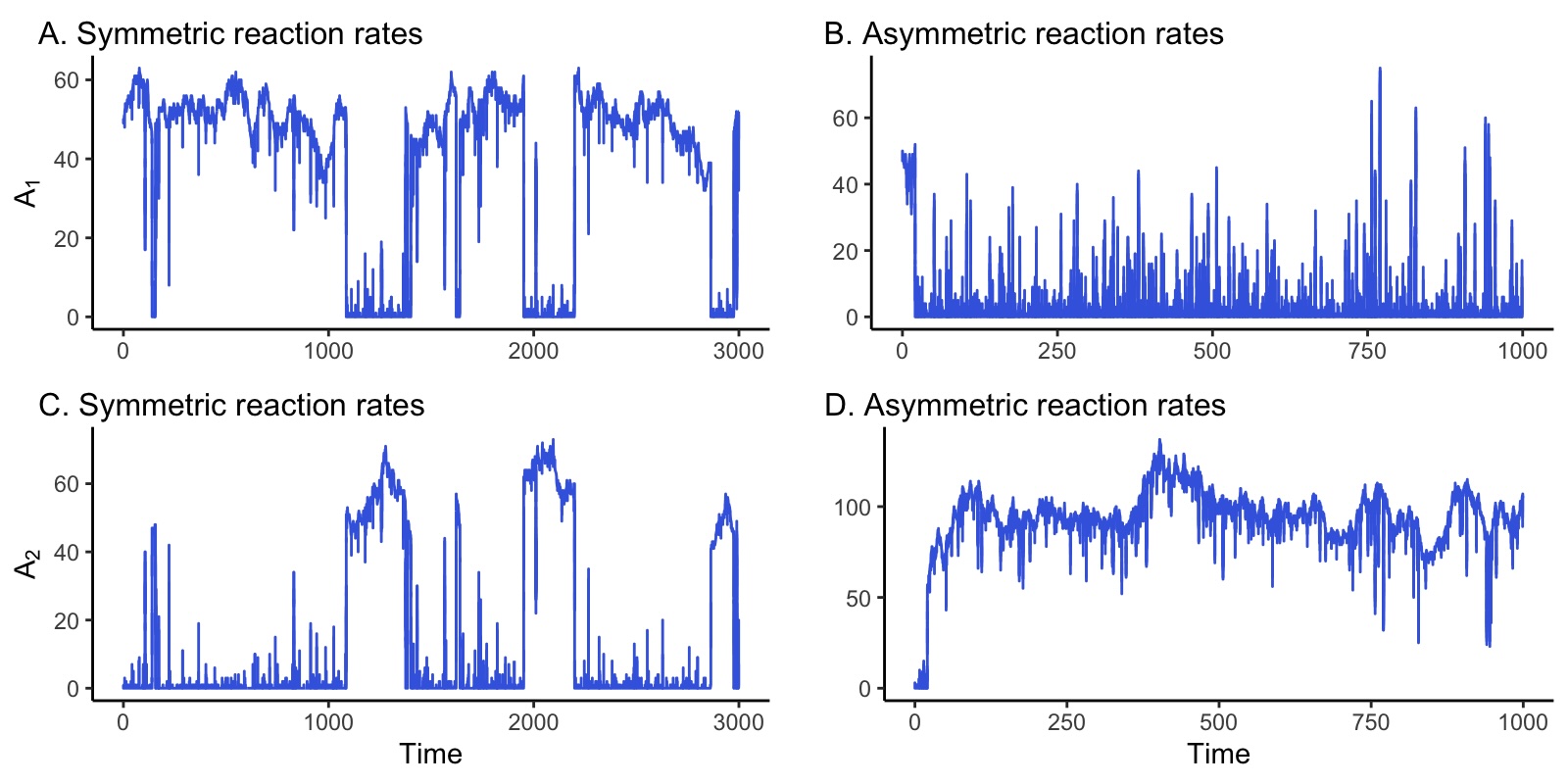}
\caption{{\bf Times series data of the TK model.} A set of representative stochastic trajectories of the TK model with initial conditions $A_1(0)=49$ and $A_2(0)=1$. 
Panels (A,C) illustrate the case of symmetric reaction rates, with $\kappa_i=200$, $\lambda_i=0.2$, and $\delta_i=0.0078$ for $i=1,2$. 
Panels (B,D) depict asymmetric rates, given by $\kappa_1=20$, $\kappa_2=19$, $\lambda_1=2$, $\lambda_2=1$, $\delta_1=0.078$, and $\delta_2=0.03$.}
\label{fig:TK}
\end{figure}

Our analysis examines two distinct scenarios of fast autocatalytic reactions: one with symmetric reaction rates and another with asymmetric rates between the two species. For each scenario, we generate 10 independent datasets, with representative examples shown in Fig~\ref{fig:TK}. Logistic regression is then applied to each dataset, using the production of species $A_2$ as the reference reaction.

In the symmetric case, stochastic trajectories of the TK model alternate between low- and high-concentration states over time (Fig~\ref{fig:TK}, left).  
By contrast, with asymmetric reaction rates, species $A_1$ fluctuates around zero while $A_2$ remains at elevated levels (Fig~\ref{fig:TK}, right).  
Although both species exhibit pronounced fluctuations in molecular counts, they do not exhibit the abrupt transitions between low- and high-concentration states that characterize the symmetric case.

From the regression results—specifically, the estimated coefficients—we can identify when the coefficients for $A_1$ or $A_2$ are statistically significant and  positive. Such significance implies that the corresponding reaction is more likely to account for the observed stoichiometric change (i.e., jump type) relative to the reference reaction. For the TK model, the production of $A_2$ serves as the reference. A significantly positive coefficient for a given set of reactants indicates that the associated reaction is more likely to occur when $A_1$, $A_2$, or both are present at elevated levels.

To better control the false positive rate, we apply a stringent criterion for identifying \emph{significant} regression coefficients, requiring one-sided $P$-values below $0.001$, which corresponds to $z$-scores of $3.09$ or higher.

For both the symmetric and asymmetric reaction rate cases in the TK model, all reactants involved in first- and second-order reactions are correctly identified, as indicated by their statistically significant positive regression coefficients and summarized in Table~\ref{table:tk} below. Further details are provided in the Supplemental Information  in Tables~\ref{table:tk_case1}--\ref{tab:tk_case2b}.

For the zeroth-order reaction, we confirm that the production of $A_1$ has no reactants in the 
symmetric reaction case (Case 1), as the coefficients for both 
species were not significant in the 
regression. In contrast, in the asymmetric reaction case with 10 
trajectories (Case 2a), the 
regression did not correctly identify the reactant for the 
production of $A_1$, as $A_1$ had a significant positive coefficient. 
However, the coefficient is very small ($0.0323$; see $A1$:4 in Table \ref{tab:tk_case2a}), corresponding to an odds ratio of approximately $1.03$,  meaning the odds of the outcome increase by approximately 3\% for each unit increase in the predictor (see, for example, \cite{Hosmer:2013:ALR}). This suggests that the productions of $ A_1 $ and $ A_2 $ are likely to occur with nearly the same frequency when $ A_1 $ is abundant. Additionally, the $ z $-value (test statistic) corresponding to the positive coefficient for $ A_1 $ in the production of $ A_1 $ is relatively modest ($ 3.915 $; see $A1$:4 in Table~\ref{tab:tk_case2a}) compared to the $ z $-values associated with other significant coefficients. This indicates that the statistical evidence supporting the influence of $ A_1 $—and, by extension, the presence of an incorrect reaction—is considerably weaker than in the other cases (see also Fig~\ref{fig:histo_tk2a} for the $z$-values distribution  pattern).

Finally, we examine the asymmetric reaction case using a dataset with a larger number of trajectories. As shown in Table~\ref{table:tk} (Case 2b), all reactants are correctly identified when analyzing 20 trajectories in total (10 from Case 2a and 10 additional ones).

\begin{table}[!ht]
\centering
\caption{
{\bf Species identification in the TK model using logistic regression.} \small
A ``+'' denotes coefficients that are both significant and positive.  
The symbols \checkmark\ and \xmark\ indicate correct and incorrect identification of the corresponding reaction, respectively. }
\begin{tabular}{|l|lll|lll|lll|}
\hline
\multicolumn{1}{|l|}{Reactions} & \multicolumn{3}{|l|}{Case 1} &
\multicolumn{3}{|l|}{Case 2a} &
\multicolumn{3}{|l|}{Case 2b}\\ \hline\hline
& $A_1$ & $A_2$& & $A_1$ & $A_2$ && $A_1$ & $A_2$ &
\\\hline
$A_1+A_2 \to 2A_1$ & + & + & \checkmark & + & + & \checkmark
& + & +& \checkmark\\ 
$A_1+A_2\to 2A_2$ & + & + & \checkmark & + & + & \checkmark
& + & + & \checkmark\\ 
$A_1\to \emptyset$ & + &  & \checkmark & + & & \checkmark
& + & & \checkmark\\ 
$\emptyset\to A_1$ & & & \checkmark & + & & \xmark
&&& \checkmark\\ 
$A_2\to \emptyset$ &  & + & \checkmark &  & + & \checkmark
& & + & \checkmark\\ \hline\hline
$\emptyset\to A_2$ (reference) & & && &&
&&&\\\hline\hline
\end{tabular}
\label{table:tk} 
\end{table}

\subsubsection*{The Heat Shock Response (HSR) model}

As our second example of network identification in an open reaction system, we consider the Heat Shock Response (HSR) model, originally developed by Linder and Rempala~\cite{Linder:2020:SLM}. This model involves two types of proteins, $ P_1 $ and $ P_2 $, as well as gene expression represented by $ R_1 $.

In the HSR model, the two proteins—heat shock transcription factors—form a positive feedback loop that promotes the production of each other. Additionally, they enhance the expression of heat shock protein genes. The gene product $ R_1 $ can self-amplify its own expression and regulate one of the transcription factors ($ P_2 $) through a negative feedback mechanism.
\begin{align*}
    \emptyset &\xrightarrow{\kappa_1} P_1, \quad \emptyset \xrightarrow{\kappa_2} P_2,  \quad \text{(Natural production)} \\
    P_2 &\xrightarrow{\kappa_3} P_1+P_2, \,\, P_1 \xrightarrow{\kappa_4} P_1+P_2, \,\, R_1 \xrightarrow{\kappa_5} 2R_1, \,\, \text{(Catalyzed production)} \\
    P_1 &\xrightarrow{\kappa_6} R_1, \quad P_2 \xrightarrow{\kappa_7} R_1, 
    \quad R_1 \xrightarrow{\kappa_8} P_2, \quad \text{(Conversion)} \\
    P_1 &\xrightarrow{\kappa_{9}} \emptyset, \quad P_2 \xrightarrow{\kappa_{10}} \emptyset, 
    \quad R_1 \xrightarrow{\kappa_{11}} \emptyset, \quad R_1+P_2 \xrightarrow{\kappa_{12}} \emptyset, \quad  \text{(Degradation)}
\end{align*}

This model features a more complex reaction network than the TK model, comprising three species and twelve reactions.  
A distinctive feature is the presence of two pairs of reactions with identical stoichiometries: the natural and catalyzed production of $P_1$ and $P_2$.  
The first pair consists of reactions with rates $\kappa_1$ and $\kappa_3$, while the second pair consists of reactions with rates $\kappa_2$ and $\kappa_4$. As a result, when reactions are classified by stoichiometry alone, both the spontaneous production of $ P_1 $ or $ P_2 $ and their catalyzed production by another species are treated as the same reaction.

In the HSR model, we assume that all reaction rate constants are of the same order of magnitude, ensuring all types of reactions occur frequently enough. We then consider two scenarios: one where all reaction rate constants are nonzero (Case 1), and another where the reaction rate constant $\kappa_2$ for the natural production of $P_2$ ($\emptyset\to P_2$) is zero (Case 2). 

\begin{figure}
\includegraphics[width=1.0\textwidth]{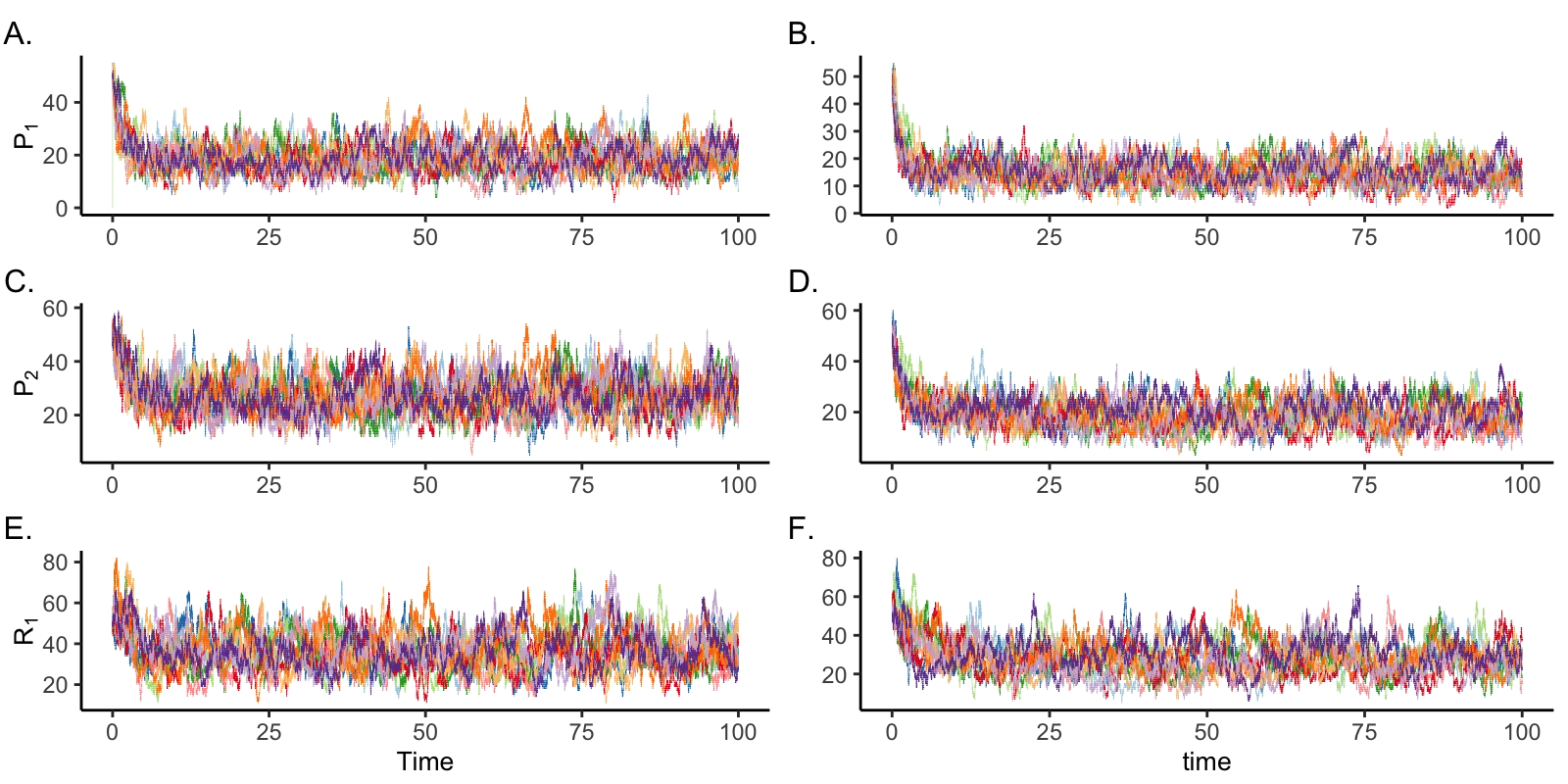}
\caption{
{\bf Time-series data of the HSR model.}  
Ten stochastic simulation trajectories of the Heat Shock Response model are shown, each initialized with $P_1(0)=P_2(0)=R_1(0)=50$.  
Panels (A,C,E) depict simulations with nonzero natural production of $P_2$, where $\kappa_1=\kappa_2=10$ and $\kappa_{12}=0.01$, with all other reaction rate coefficients set to $1$.  
Panels (B,D,F) correspond to simulations with no natural production of $P_2$ ($\kappa_2=0$), while all other reaction rate coefficients remain identical to those in Panels (A,C,E).
}
\label{fig:HS}
\end{figure}

Fig~\ref{fig:HS} presents ten simulated trajectories for each case.
The left panels show the trajectories of three species in Case 1, while the right panels depict those for Case 2. In Case 2, the overall levels of $P_1$ and $P_2$ are slightly reduced compared to Case 1, due to the absence of natural production of $P_2$.

In Case 1, we set the production of $P_2$ as the reference reaction and perform logistic regression as in the TK model. In this case (results shown in Table~\ref{table:heat_shock}), all reactants involved in the chemical reactions are correctly identified using $10$ simulated trajectories, even though the reference reaction includes both the natural and catalyzed production of $P_2$. 

In Case 2, where natural production of \(P_2\) is absent (\(\kappa_2 = 0\)), the reference reaction consists solely of the catalyzed production of \(P_2\), which uses \(P_1\) as a reactant. Because this reference reaction shares both the same reactant set (\(P_1\)) and identical reaction rates with two other reactions, the logistic model encounters an identifiability issue: the substrate abundance is identical across three reactions, making it impossible to distinguish the reaction type from \(P_1\) alone. This scenario is intentionally constructed to illustrate how the choice of reference reaction influences the ability to identify the underlying chemical network from trajectory data. For this case, we first analyze a dataset of ten simulated trajectories (Case~2a) and then extend the dataset by adding ten additional trajectories (Case~2b).

As seen in Table~\ref{table:heat_shock} in both scenarios denoted as Case 2a and Case 2b, $P_1$ is not identified as a reactant in the conversion and degradation reactions, as expected.
Despite this, the logistic regression correctly identifies the reactants in all remaining reactions in both cases, as summarized in Table~\ref{table:heat_shock}. Further details are given in the Supplemental Material in Tables~\ref{table:heat_shock_case1}-\ref{tab:hs_case2b}. 

In addition to identifying significant positive coefficients for the reactants in Table~\ref{table:heat_shock}, the distribution of the  $z$-values for all species across all reactions is summarized in Fig~\ref{fig:histo_hs2a} (Case 2a) and Fig~\ref{fig:histo_hs2b} (Case 2b). In these histograms, the red vertical line marks the threshold $z$-value of $3.09$, corresponding to a $P$-value of $0.001$. Orange bars to the right of this line represent significant positive coefficients ($P<0.001$).The distribution of $z$-values is clearly separated around this threshold, and the separation becomes more pronounced as the dataset size increases from Case 2a to Case 2b (see Figs~\ref{fig:histo_hs2a}–\ref{fig:histo_hs2b}). This trend highlights that larger time-series datasets enhance the reliability of network structure identification.

\begin{table}[!h]
\centering
\caption{
{\bf Species identification in the HSR model using logistic regression.} \small
A ``+'' indicates that the estimated coefficients are significant and positive.  
The symbol \checkmark\ denotes correct identification of the corresponding reaction. In Case~2, the absence of natural \(P_2\) production (\(\kappa_2 = 0\)) creates an identifiability issue, as the reference reaction shares its sole reactant \(P_1\) and reaction rates with two other reactions. This makes the reaction types indistinguishable from \(P_1\) alone and is reflected by the blanks in rows~1 and~7 of the table, which correspond to reactions involving \(P_1\).
}
\begin{tabular}{|l|llll|llll|llll|}
\hline
\multicolumn{1}{|l|}{ Reactions} & 
\multicolumn{4}{|l|}{Case 1} &
\multicolumn{4}{|l|}{Case 2a} &
\multicolumn{4}{|l|}{Case 2b}\\ \hline\hline 
& $P_1$ & $P_2$ & $R_1$ && $P_1$ & $P_2$ & $R_1$ && $P_1$ & $P_2$ & $R_1$ &\\\hline
$P_1 \to R_1$ & + &  &  & \checkmark &  &  &  & 
&&&&\\ 
$P_2\to R_1$ &  & + &  & \checkmark &  & + & & \checkmark
& & + & & \checkmark\\ 
$R_1\to P_2$ &  &  & + & \checkmark &  &  & + & \checkmark
& & & + & \checkmark\\ 
$R_1\to 2R_1$ &  &  & + & \checkmark &  &  & + & \checkmark
& &  & + & \checkmark\\ 
$R_1+P_2\to \emptyset$ &  & + & + & \checkmark &  & + & + & \checkmark
& & + & + & \checkmark\\ 
$R_1 \to \emptyset$ & & & + & \checkmark & & & + & \checkmark
& & & + & \checkmark\\ 
$P_1\to \emptyset$ & + & & & \checkmark & & & &
&&&&\\ 
$P_2\to \emptyset$ & & + & & \checkmark & & + & & \checkmark
& & + & & \checkmark\\ 
$\left\{\begin{array}{l}
\emptyset\to P_1,\\
P_2\to P_1+P_2
\end{array}\right.$
& & + & & \checkmark & & + & & \checkmark
& & + & & \checkmark\\ 
\,\,\,\,
& & & && & & &
&&&&\\\hline\hline
\multicolumn{1}{|l|}{ 
$\left\{\begin{array}{l}
\emptyset\to P_2,\\
P_1\to P_1+P_2
\end{array}\right.$ 
}
& 
\multicolumn{4}{|c|}{} 
&
\multicolumn{8}{|c|}{ 
$\emptyset\to P_2$ 
}
\\
\multicolumn{1}{|l|}{\,\,\,\,(reference)}
& 
\multicolumn{4}{|c|}{} 
&
\multicolumn{8}{|c|}{ 
removed 
}
\\
\hline\hline
\end{tabular}
\label{table:heat_shock}
\end{table}

\subsubsection*{SIR model with demography}
As a final example—used here to illustrate both structural and parameter inference—we consider the classical Susceptible–Infected–Recovered (SIR) reaction network with demographic turnover. This model is widely used for studying epidemic dynamics~\cite{Kermack:1927:CMT,Britton:2010:SEM}. It includes three types of species: susceptible (\( S \)), infected (\( I \)), and recovered (\( R \)) individuals in a well-mixed population. The model describes the transmission process whereby susceptible individuals become infected through contact with infected individuals, and later transition to the recovered class.

The version of the SIR model considered here incorporates demographic effects by accounting for natural birth and death processes (or, equivalently, immigration and emigration) for all species. The resulting reaction network accounts for the natural turnover of the population and is given by
\begin{align*}
    S + I &\xrightarrow{\beta/n} 2I, \quad \text{(Disease transmission)} \\
    I &\xrightarrow{\gamma} R, \quad \text{(Recovery process)} \\
    \emptyset &\xrightarrow{\mu n} S, \quad \text{(Natural birth/immigration)} \\
    S &\xrightarrow{\nu} \emptyset, \quad I \xrightarrow{\nu} \emptyset, \quad R \xrightarrow{\nu} \emptyset. \quad \text{(Natural death/emigration)}
\end{align*}
Here, the parameter $n$ serves as a maximal system size over a pre-specified time window, or more generally, as a scaling parameter that determines the scale at which the population dynamics are considered.
Despite its relative simplicity, this model has been successfully applied and analyzed in the study of several global epidemics—including by some of the present authors—in contexts such as H1N1~\cite{Schwartz:2015:EEP,Khudabukhsh:2020:SDS}, Ebola~\cite{Wairimu:2024:PNS}, COVID-19~\cite{Linder:2022:EMP}, and other infectious diseases.

As in the previous sections, we generate synthetic trajectories of the network and apply a logistic regression approach to identify the reactant species involved in each reaction within the extended SIR model. The inflow  of susceptibles (representing natural birth or immigration) is used as the reference reaction.
\begin{figure}
\includegraphics[width=1.0\textwidth]{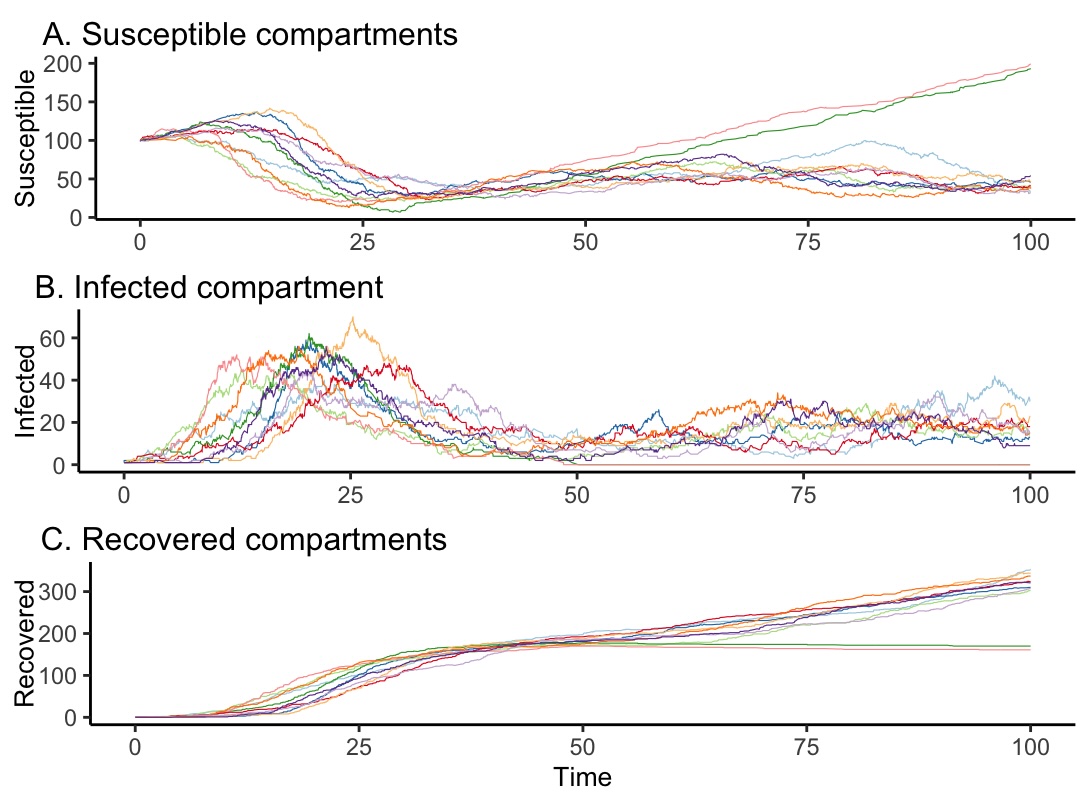}
\caption{{\bf Time series data from the SIR model with demography.} Ten stochastic simulation trajectories of the SIR model are shown, selected based on sustained infection spread (large outbreak). All runs begin with initial conditions $S(0) = 100$, $I(0) = 1$, and $R(0) = 0$. Reaction rate coefficients are: $\beta/n = 0.004$, $\gamma = 0.2$,  $\mu n=3$, and $\nu=0.001$.}
\label{fig:SIR}
\end{figure}
Fig~\ref{fig:SIR} illustrates the evolution of the three species counts over time across the 10 selected trajectories. Due to the continuous inflow of susceptible individuals, the overall dynamics differ slightly from those of the basic SIR model (i.e., one defined on a closed population without demography). In particular, both the susceptible and recovered populations exhibit a gradual increase toward the end of the observation period. In contrast, the temporal evolution of the infected population—reflecting epidemic incidence—closely resembles the dynamics observed in the basic SIR model.

The results based on the logistic regression model and data from 10 simulated trajectories are summarized in Table~\ref{table:sir} with further details provided in the Supplemental Information (Tables~\ref{table:sir2} and \ref{tab:sir3} ). As shown in Table~\ref{table:sir}, the reactants involved in all reactions are correctly identified. Most of the estimated regression coefficients are positive; however, one notable exception is a significantly negative coefficient for the susceptible species in the recovery reaction (see S:2 in Table~\ref{tab:sir3}). This negative value can be interpreted as indicating a relative preference among competing reactions. Specifically, as the number of susceptibles increases, the birth reaction becomes more likely than the recovery reaction. This interpretation aligns with the core rationale of logistic regression, which is to identify species whose presence either increases or decreases the likelihood of a given reaction occurring, relative to a specified reference. The significantly negative coefficient may therefore be viewed as strong evidence against including $S$ as a reactant in the recovery reaction.

\begin{table}[!ht]
\centering
\caption{
{\bf Species identification in the SIR model with demography using logistic regression.} \small A ``+'' indicates that the estimated coefficients are significant and positive.  
The symbol \checkmark\ denotes correct identification of the corresponding reaction. }
\begin{tabular}{|l|llll|}
\hline
\multicolumn{1}{|l|}{Reactions} & \multicolumn{4}{|l|}{Species}\\ \hline\hline
& $S$ & $I$ & $R$ &\\\hline
$S+I \to 2I$ & + & + &  & \checkmark\\ 
$I\to R$ &  & + & & \checkmark\\
$S\to \emptyset$ & + &  & & \checkmark\\ 
$I\to \emptyset$ &  & + & & \checkmark\\ 
$R\to \emptyset$ &  &  & + & \checkmark\\ \hline\hline
$\emptyset\to S$ (reference) & & & &\\\hline\hline
\end{tabular}

\label{table:sir}
\end{table}

\section*{Parameter estimation}
After identifying the reactions in a chemical network, estimating the associated parameters is essential for understanding system dynamics and making reliable predictions.  
While statistical challenges such as stochasticity and partial observation arise in most biochemical systems, epidemic models offer a more tractable setting because population-level data are often readily available.  
For that reason  we focus on the SIR  epidemic model discussed above and show how logistic regression can also be employed for parameter estimation using real outbreak data.

\subsection*{Case study: COVID-19 data analysis}


Having successfully identified the network structure from synthetic data, we now turn to parameter estimation using real-world observations. In this section, we propose a novel method for estimating the parameters of the SIR model with demography, grounded in the network identification framework based on logistic regression. We apply this method to epidemic data from the COVID-19 outbreak in Seoul, South Korea discussed already in different context in ~\cite{Hong:2024:OBE}.

The dataset includes the time of symptom onset and the time of confirmation for each individual case. Since an infected individual is typically capable of transmitting the virus shortly after symptom onset, we identify the time of infectiousness onset as the time of infection~\cite{Hong:2024:OBE}. Upon confirmation of infection, individuals are immediately isolated; thus, we take the date of confirmation to represent the time of removal, as the individual is no longer contributing to transmission.

\begin{figure}[t!]
\includegraphics[width=1\linewidth]{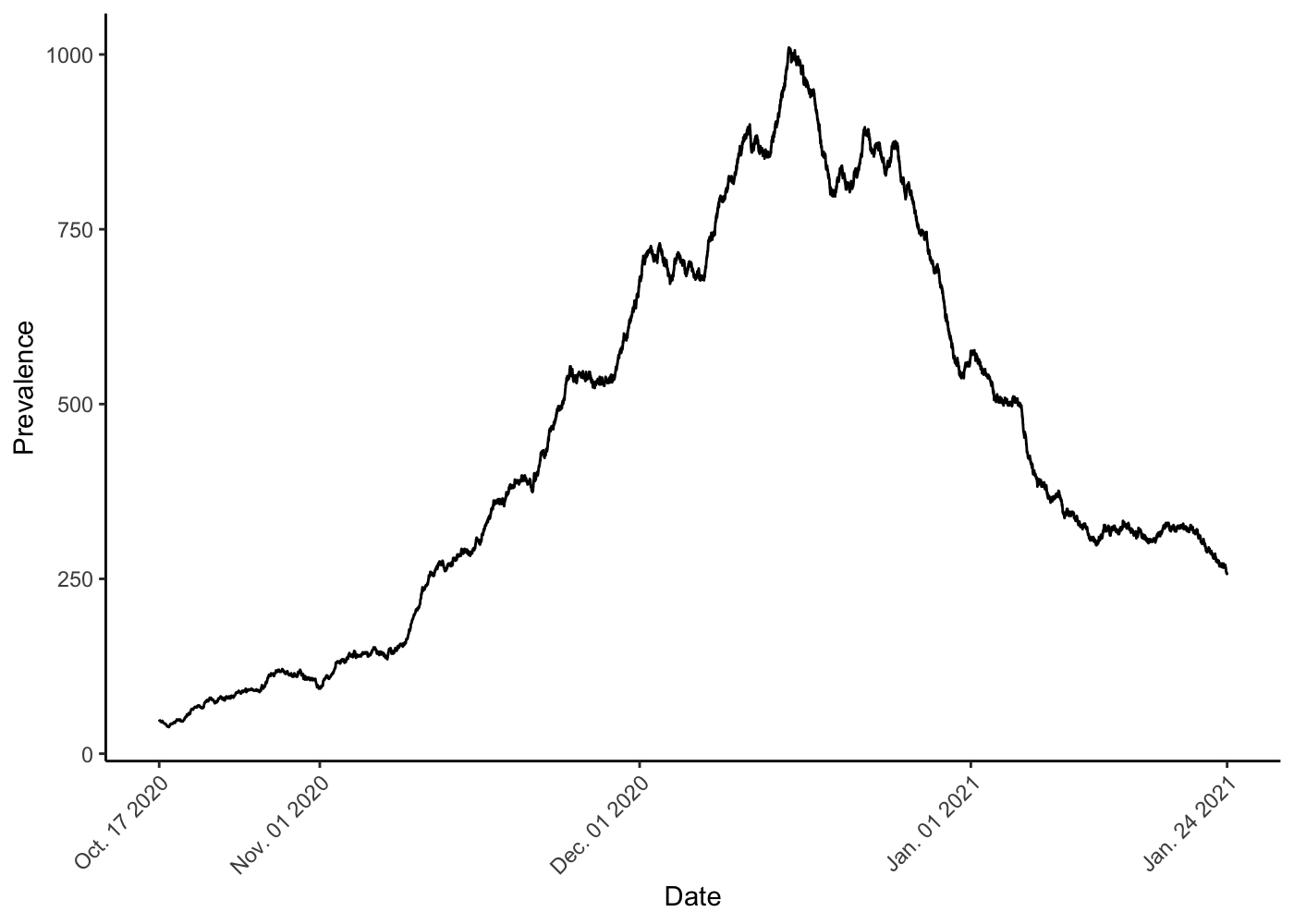}
\caption{\textbf{Prevalence of COVID-19 of Seoul, South Korea from Oct. 17, 2020, to Jan. 24, 2021.}}
\label{fig:prevalence}
\end{figure}

Using this information, we construct a time-series trajectory of the infected population. Figure~\ref{fig:prevalence} illustrates the prevalence from October 17, 2020, to January 24, 2021.

Accordingly, the infection process is modeled by the following stochastic equation \cite{Choi:2011:IDO}.
 
\begin{align}
    I_t &= I_0 + Y_1 \left( \int_0^t \frac{\beta}{n} S_u I_u\, \differential{u} \right) - Y_2 \left( \int_0^t ({\nu}+\gamma) I_u\, \differential{u} \right), \label{i_popul}
\end{align}
where $I_t$ is the count of infected at time $t$, $Y_1$ and $Y_2$ are independent unit Poisson processes and the parameter  $n$ represents the {\em effective population size}.
Note that the recovery and degradation of the infected population are represented as a single term in \eqref{i_popul}, since they follow the same distribution as the infected population in the SIR model. 
The infected population at each time point is denoted as $\tilde{I}=(i_1,i_2,\cdots,i_\ell)$, with corresponding observation times $\tilde{T}=(t_1,t_2,\cdots,t_\ell)$, where $t_j$ is the $j$-th observation time, and we assume all infection events are recorded.
We define the infection event indicator as $\tilde{Y}=(y_1,y_2,\cdots,y_{\ell-1})$, where 
\begin{equation}
y_j = \left\{
\begin{array}{ll}
1 & \mbox{if $i_{j+1}-i_j=1$},\\
0 & \mbox{if $i_{j+1}-i_j=-1$},\\
\end{array}
\right.
\end{equation}
for $j=1,2,\cdots,\ell-1$.
In this context, $y_j=1$ corresponds to an infection event, and $y_j=0$ corresponds to a recovery event. So $\tilde{Y}$ can be the observed response data for the parameter estimation using the logistic regression model. 

We can construct the likelihood using the observed response data $\title{Y}$ and apply the logistic regression model approach described in the Methods section.   At each time $t$, the ratio of infection event probability ($p_t$) to recovery event probability ($1-p_t$) is given by
\begin{equation}
\frac{p_t}{1-p_t} = \frac{\frac{\beta}{n} S_t \,I_t}{\gamma I_t}
= \frac{\beta}{\gamma} \frac{S_t}{n}.\label{i_ratio} 
\end{equation}
Taking the logarithm of \eqref{i_ratio}, we obtain the log-odds as
\begin{eqnarray}
\log{\left(\frac{p_t}{1-p_t}\right)} &=& \log{\left(\frac{\beta}{\gamma}\right)} + \log{\left(\frac{S_t}{n}\right)}. 
\label{log_odds}
\end{eqnarray}

Since direct observation of $S_t/n$ is not feasible, we approximate it using the law of large numbers relation (see, for instance \cite{Anderson:2015:SAB}) $S_t/n\approx s_t$, where $s_t$ is governed by the deterministic SIR model with birth and death processes:
\begin{align}\label{eq:demsir}
    \dot{s}_t &= -\beta s_t \iota_t +\mu-\nu s_t , \nonumber \\
    \dot{\iota}_t &= \beta s_t \iota_t - \gamma \iota_t - \nu \iota_t, \\
    \dot{r}_t &= \gamma \iota_t - \nu r_t.\nonumber
\end{align}
The initial conditions are  $s_0=1$, $\iota_0=\rho$, and $r_0=0$.
Given that migration rates in Seoul exceed natural birth and death rates, we model the combined effects of immigration and natural births as a single effective birth rate (\( \mu \)), and similarly combine emigration and natural deaths into an effective death rate (\( \nu \)). These demographic rates are calibrated using the net population movement rate during the observation period, which is approximately (based  on the census data) \( 0.013\,\mbox{day}^{-1} \). Assuming that the total population of Seoul is approximately $10^7$, and only a fraction $n$ of the population is susceptible and may participate in infection events, we estimate the effective  birth and death rates as
\begin{eqnarray*}
\mu_n = 0.013\times  \frac{n}{10^7},\quad \nu_n = \frac{0.013}{3} \times \frac{n}{10^7}.
\end{eqnarray*}

Note that \( n \) denotes the \emph{effective population size}, which is one of the quantities to be estimated and must therefore be updated during the MCMC procedure. As \( n \) evolves throughout the MCMC iterations, the associated birth and death rates, \( \mu_n \) and \( \nu_n \), also vary accordingly. To simplify the inference process, we fix the effective population size at a chosen time horizon \( T > 0 \), during which we observe two types of events: infections and recoveries.

Given the total number of observed infections, we approximate the numerical value of the parameter $n$ as the mean of a random variable $N$ drawn from a negative binomial distribution
\begin{equation}\label{effective_popul}
N \sim \mathrm{NegBinomial}(n_I, \tau_T),
\end{equation}
where $n_I$ denotes the total number of observed infections, and $\tau_T$ represents the probability that a susceptible individual becomes infected by time $T$, accounting for demographic events. The value of $n_I$ is derived from real epidemic data—for example, from observed COVID-19 outbreaks—and is computed as $\sum_{j=1}^{\ell-1} y_j$, representing the cumulative number of infections up to time $T$.

The probability $\tau_T$ is calculated (see \eqref{eq:tau}) as:

$$
\tau_T = \frac{1 - s_T + \mu T - \int_0^T \nu s_u \, \differential{u}}{1 + \mu T - \int_0^T \nu s_u \, \differential{u}}.
$$
This definition of $\tau_T$ ensures that \eqref{effective_popul} yields an estimate of the effective population size that adjusts for birth and death events over the observation window.

According to the log-odds of infection occurrence given in (\ref{log_odds}), we may formulate a logistic regression model as:

\begin{equation} \label{logistic_model}
    \log\left(\frac{P(Y = 1)}{1 - P(Y = 1)}\right) = \alpha + \mathrm{offset}(\log s_t),
\end{equation}
where $Y$ is a binary indicator of an observed infection event, $s_t$ is the susceptible fraction obtained from the solution of the ODE system (\ref{eq:demsir}), and $\alpha$ is the intercept of the logistic regression model.

We define the vector of unknown parameters to be estimated as 
\begin{equation}\label{eq:theta} \theta = (\beta, \gamma, \rho),\end{equation} and the corresponding likelihood function is given by (see \eqref{eq:lkhd}):
\begin{eqnarray*}
L (\theta,n) = \prod_{j=1}^{\ell-1}  P(Y=1)^{y_j}(1-P(Y=1))^{(1-y_j)}.
\end{eqnarray*}
Note that when $Y=y_j$, the concentration of the susceptible population corresponds to $s_t=s_{t_j}$ for $j=1,2,\cdots,\ell-1$.
The prior distributions for the parameters vector $\theta$ in \eqref{eq:theta} are defined independently as 
\begin{equation}
\begin{split}
& \beta \sim \mbox{Gamma}(10^{-3},10^{-3})\equiv f(\beta),\\
& \gamma \sim \mbox{Gamma}(0.25\times 10^4,10^4)\equiv g(\gamma),\\
& \rho\sim \mbox{Beta}(1,1)\equiv h(\rho). 
\end{split}
\label{priors}
\end{equation}
Since we have the time of infection and and time of recovery, we can directly estimate the infectious period. So we assigned $\gamma$ an informative prior using the information about the mean infectious period.  According to \eqref{eq:propto}
the posterior distribution  then satisfies
\begin{eqnarray}
q(\theta,n) &\propto & L(\theta,n)\, f(\beta)\,g(\gamma)\,h(\rho). 
\label{posterior}
\end{eqnarray}
Finally, since the posterior distribution in~\eqref{posterior} is complex and lacks a closed-form expression, we employ Markov Chain Monte Carlo (MCMC) methods discussed earlier to perform Bayesian estimation of the parameters \( \theta \). Specifically, we use the Metropolis–Hastings algorithm within a Gibbs sampler~\cite{Wilkinson:2018:SMS}, incorporating an offset computed from the solution of the deterministic SIR model with birth and death processes given in~\eqref{eq:demsir}. The proposed MCMC procedure is outlined in Algorithm~\ref{alg:mcmc}.

Using this algorithm, we run 4,000 iterations and remove the first half of the iterations as a burn-in set. The last 2000 iterations of the Metropolis–Hastings sampler are used to estimate all parameters, including the effective population size \( n \). As shown in Fig~\ref{fig:trace}, the MCMC simulation quickly reaches stationarity, and all four parameters exhibit good mixing behavior.

\begin{algorithm}
\caption{MCMC algorithm for Bayesian logistic regression}
\label{alg:mcmc}
\begin{algorithmic}[1]
\STATE Initialize all parameters $(\theta,n)$ based on the prior distributions in \eqref{priors}. Initialize the effective population size as $n=3\times 10^4$. 
\STATE Solve the system of ODEs in \eqref{eq:demsir} using the current values of $(\theta,n)$ to obtain $s_t$. 

\STATE Draw samples of $\theta$ from their posterior \eqref{posterior} using the Robust Adaptive Metropolis (RAM) algorithm~\cite{Vihola:2012:RAM}. 
Update the intercept $\alpha$ in the logistic model (\ref{logistic_model}). 

\STATE Update the effective population size $n$ sampled from the negative binomial distribution (\ref{effective_popul}) and $n_I$.

 \STATE Repeat steps 2-4 until convergence.
\end{algorithmic}
\end{algorithm}


Table~\ref{tab:summary_stats} summarizes the MCMC simulation results for the Bayesian inference. For all parameters, the posterior means and medians are similar, indicating that the posterior distributions are approximately symmetric and that the credible intervals are relatively narrow. Additional diagnostics for the MCMC simulation are provided in the appendix. Fig~\ref{fig:trace} shows the trace plots for the posterior samples of the parameters, demonstrating rapid convergence to stationarity. Fig~\ref{fig:corr} displays histograms of the posterior samples for each parameter, along with paired scatter plots. With the exception of \(\rho\), all histograms are approximately symmetric. Some linear relationships are evident between \(\beta\) and \(\gamma\), as well as between \(\beta\) and the effective population size, but these remain within acceptable limits.

The posterior mean of the basic reproduction number, $R_0=\beta/\gamma$, is estimated to be $1.292$. This estimate closely aligns with previous findings for Seoul~\cite{Yoo:2021:CRN}. The last column of Table~\ref{tab:summary_stats} summarizes the estimates of the effective population size, $n$, which is approximately $30,000$ during the observation period in Seoul. 

\begin{figure}
\includegraphics[width=1.0\textwidth]{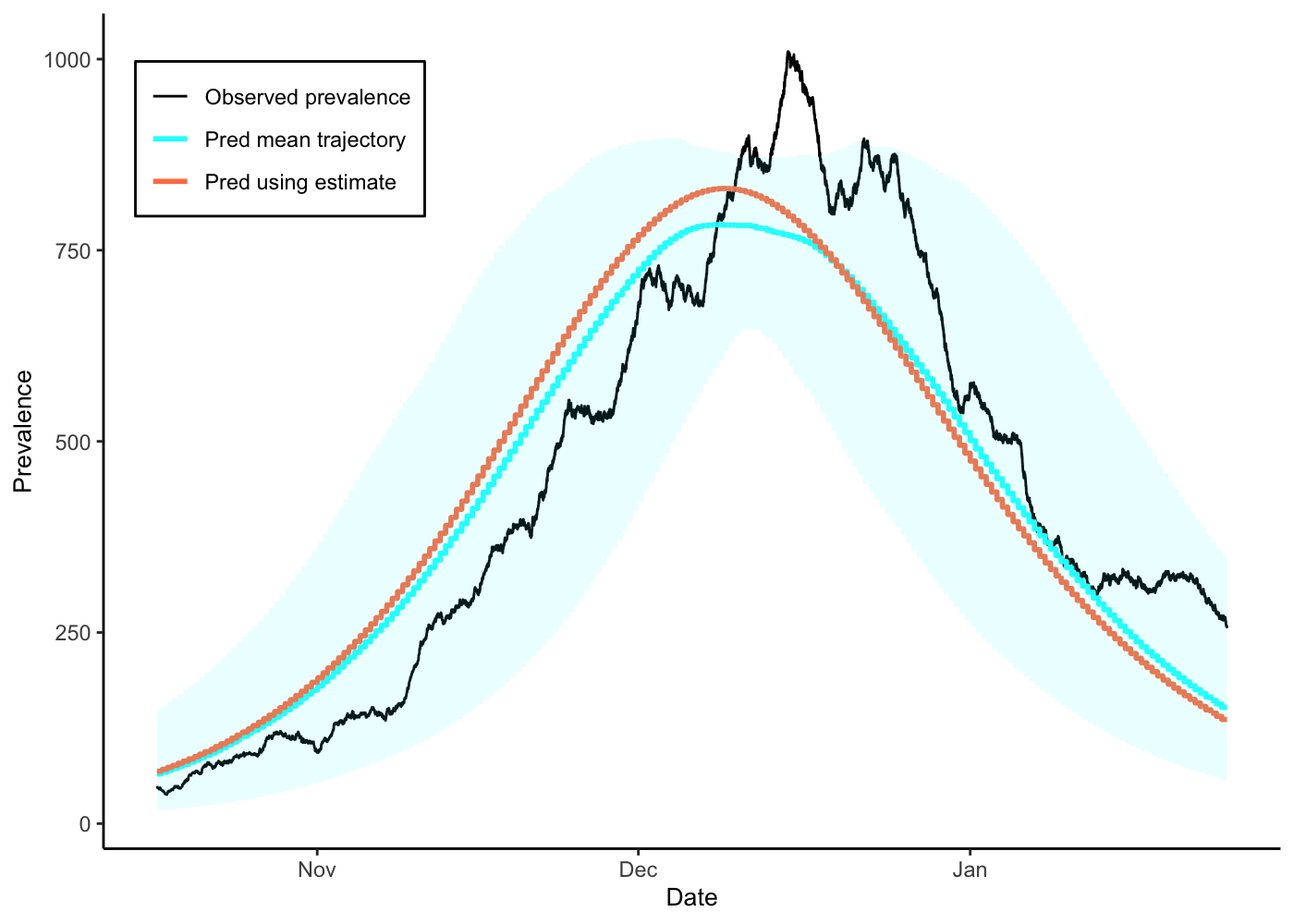}
\caption{
{\bf Comparison of observed and estimated infection prevalence.} \small
The black line shows the observed number of infected individuals over time. The blue line represents the estimated mean infection prevalence, computed as the average of trajectories obtained by solving the SIR model using posterior samples of \(\beta\), \(\gamma\), \(\rho\), and \(n\). The red line shows the infection curve generated by solving the SIR model with demography, using the posterior means of the estimated parameters.
}
\label{fig:prediction}
\end{figure}

In Fig~\ref{fig:prediction}, the observed prevalence, the number of infected individuals (black line) is compared to the estimated mean number (blue line) obtained from the Bayesian estimation results. This curve is the mean of the trajectories by solving the SIR model using the set of posterior samples of $\beta$, $\gamma$, $\rho$, and $n$. The red line represents the estimated infection curve, which was generated by solving the SIR model with birth and death processes, using the posterior means of the estimated parameters: $\beta$, $\gamma$, $\rho$, and $n$. Overall, the estimated trend closely follows the observed values. The shaded area represents the 95\% credible band for the fitted prevalence curve, which covers most of the observed prevalence. 
The sharp rise in infections during the early phase of the outbreak is well captured by the model. However, the estimated peak occurs slightly earlier than the actual peak observed in the data.


\begin{table}[htbp]
\centering
\begin{tabular}{lrrrc}
\hline
Statistic & $\beta$ & $\gamma$ & $\rho$ & Eff. pop. size $n$\\
\hline
Min.     & 0.2993 & 0.2351 & 0.000349 & 23323 \\
1st Qu.  & 0.3159 & 0.2468 & 0.001570 & 26276 \\
Median   & 0.3217 & 0.2495 & 0.002297 & 27400 \\
Mean     & 0.3221 & 0.2499 & 0.002401 & 27570 \\
3rd Qu.  & 0.3282 & 0.2535 & 0.003085 & 28751 \\
Max.     & 0.3498 & 0.2689 & 0.007946 & 33145 \\
\hline
\end{tabular}
\caption{Summary statistics for the parameters $\beta$, $\gamma$, $\rho$, and the effective population size (number of individual at risk of exposure) $n$.}
\label{tab:summary_stats}
\end{table}

\section*{Summary and discussion}

Assuming that all chemical reactions follow the law of mass action in a well-mixed environment, we modeled a stochastic chemical reaction network as a continuous-time Markov jump process. Constructing such a model presents two primary challenges: identifying the underlying network structure and inferring its parameters.

To address these challenges, we developed a novel, to the best of our knowledge, likelihood-based approach using logistic regression. This approach is particularly well-suited to open networks that include birth reactions, where species can enter the system from an external source. When full time-series data are available—including molecular counts and reaction events for all species—logistic regression can effectively identify the network structure, specifically the set of reactants involved in each reaction. By designating a production reaction as the reference category, we applied the logistic regression model to the observed reaction events. Species with statistically significant positive coefficients were identified as reactants, as increases in their molecular counts raise the probability of the corresponding reaction occurring.

The reliability of reactant identification is sensitive to the choice of significance thresholds (e.g., \(P\)-value cutoffs), which can strongly influence accuracy. Because the logistic regression model involves a large number of potential coefficients, the chosen cutoff typically needs to be more stringent than the conventional significance level. In practice, a good cutoff value may often be determined by visually inspecting the distribution of \(P\)-values or \(z\)-values; in our analysis, this pragmatic approach proved effective for balancing stringency with interpretability. Building on the identified set of reactants, we may then construct the propensity functions for the continuous-time Markov chain model using the law of mass action and estimate the corresponding model parameters through an additional logistic regression step. This stage also allows us to account for structural imperfections in the data, such as partially missing observations or unmeasured variables, ensuring a more robust parameter estimation process.  

To illustrate the application of the logistic regression model for parameter estimation in the presence of partially missing reaction network data, we applied this approach to real-world time-series data on COVID-19 epidemics. The dataset consisted of infection prevalence measured over a fixed observation period. We assumed that the data followed an expanded SIR reaction network that accounted for population birth and death (or migration) processes. Because only the infected population was observed, with no direct measurements of susceptible or recovered individuals, we modified the logistic regression framework by incorporating an offset term. This offset captured the influence of the unobserved species and was approximated using a deterministic SIR model—the mean-field limiting ODE system derived from the stochastic SIR model via the law of large numbers.

Based on the offset-modified logistic regression model, we developed an algorithm for Bayesian logistic regression analysis and applied it to the stochastic SIR model, in which model parameters were estimated through iterative steps combining MCMC sampling with numerical solutions of the corresponding ODE system. This algorithm enabled joint estimation of key epidemiological parameters, including infection and recovery rates, the initial proportion of infected individuals, and the effective population size.

A subtle challenge arose from the interdependence between the effective population size and the ODE solutions, which in turn affected the offset term in the logistic regression. To address this, we developed an algorithm that integrates MCMC sampling with ODE solving, enabling the parameter estimates to evolve through iterative refinement. This strategy was seen to yield a robust and flexible framework for inference in partially observed epidemic systems governed by stochastic dynamics.

Despite the relative effectiveness of our logistic regression and likelihood-based methods for network identification and parameter estimation, several limitations remain, particularly in settings with incomplete or partially observed data. The network identification procedure assumes full observability of stoichiometry and time-series molecular data, which is often unrealistic, while the parameter estimation method, though accommodating partial observations such as infection and recovery times, still depends on external deterministic approximations to construct offset terms in the logistic regression. Without such supporting models, inference can become unreliable or statistically unidentifiable. Extending this framework to handle unobserved species and incomplete time-series data is therefore a critical direction for future research, enabling broader application of these techniques to real-world systems where full observability is rarely achievable.

Taken together, our results highlight the potential of combining logistic regression with likelihood-based inference as a flexible strategy for analyzing complex stochastic systems, even in the presence of intrinsic noise and incomplete data. By extending these methods to better accommodate unobserved components and irregular observations, this framework could become a powerful tool for uncovering mechanistic structure and dynamics across diverse domains—from genetic and cellular networks to ecological and social systems—where rigorous, data-driven inference remains a fundamental challenge.





\section*{Acknowledgments}
The authors gratefully acknowledge the organizers of the \textit{Workshop on the Chemical Reaction Network Theory} held at POSTECH (Pohang University of Science and Technology) in the Republic of Korea in July 2024, where this project was initiated.

\section*{Data and code availability} 

The simulated time-series datasets for the TK, HSR, and SIR models generated in this study have been deposited in a publicly accessible GitHub repository ().
All analyses were performed using R (version 4.4.2), and all code developed by the authors is publicly available on GitHub ().

\section*{Conflicts of interest} The authors declare no conflicts of interest. 

\section*{Human subjects} Real-world COVID-19 prevalence data were collected with informed consent and provided by the Seoul Metro Infectious Disease Research Center. The Korea Public Institutional Review Board, designated by the Ministry of Health and Welfare, waived the requirement for ethical approval for the collection and analysis of these data, as all information was anonymized and no individuals were identifiable (reference number: P01-202404-01-016).

\section*{Author contributions}
HWK: Writing – original draft preparation; formal analysis.\\
\noindent BC: Formal analysis; writing – original draft preparation; software (R scripts for data processing and visualizations).\\
\noindent GAR (corresponding author): Conceptualization; methodology; writing – original draft preparation.\\
\noindent All authors: Writing – review \& editing; validation; approval of the final manuscript.

\nolinenumbers

%
%
%

\bibliography{logistic}

\clearpage
\renewcommand{\normalsize}{\small}

\renewcommand{\thefigure}{S\arabic{figure}}
\setcounter{figure}{0}

\renewcommand{\theequation}{S\arabic{equation}}
\setcounter{equation}{0}

\section*{Supplemental Material: Tables and Figures}
\subsection*{Tables}
\renewcommand{\thetable}{S\arabic{table}}
\setcounter{table}{0} 

\begin{table}[!ht]
\centering
\caption{{\bf Species identification in the TK model using logistic regression (Case 1).}\small  \   Symbol ``+'' indicates that estimated coefficients are significant and have positive signs. Symbol
\checkmark\ indicates correct identification of the corresponding reaction.}
\begin{tabular}{|c|l|lll|r|}
\hline
No. & \multicolumn{1}{c|}{Reaction} & $A_1$ & $A_2$ &  & \# of Obs. \\\hline\hline
1 & $A_1+A_2 \to 2A_1$             & + & + & \checkmark & 173,174\\
2 & $A_1+A_2 \to 2A_2$             & + & + & \checkmark & 172,751\\
3 & $A_1 \to \emptyset$            & + &   & \checkmark & 6,535\\
4 & $\emptyset \to A_1$            &   &   & \checkmark & 5,982\\
5 & $A_2 \to \emptyset$            &   & + & \checkmark & 5,495\\
6 & $\emptyset \to A_2$ (reference)&   &   &            & 6,062\\\hline\hline
\end{tabular}
\begin{flushleft}

\end{flushleft}
\label{table:tk_case1} 
\caption{{\bf Multinomial logistic regression model fitting summary table for the TK model with symmetric reaction rate. (Case 1)} \small \ Highly significant positive values are shown in bold. Refer also to the histogram of $z$-values in Fig~\ref{fig:histo_tk1}.}
\begin{tabular}{crrr}
\hline Reactant:Reaction
 & Estimate & Std. Error & $Z$-value \\
\hline
$A_1$:1 & 0.0492 & 0.0020  & $~~{\bf 24.575}^{***}$ \\
$A_1$:2 & 0.0492 & 0.0020  & $~~{\bf 24.594}^{***}$ \\
$A_1$:3 & 0.0721 & 0.0034 & $~~{\bf 21.346}^{***}$ \\
$A_1$:4 & 0.0039 & 0.0028  & $~~~~1.412~~~$ \\
$A_1$:5 & -4.8569 & 0.1567 & $-30.991^{***}$ \\
$A_2$:1 & 0.0592 & 0.0021  & $~~{\bf 28.522}^{***}$ \\
$A_2$:2 & 0.0588 & 0.0021  & $~~{\bf 28.332}^{***}$ \\
$A_2$:3 & -6.6513 & 0.3538 & $-18.801^{***}$ \\
$A_2$:4 & 0.0049 & 0.0029  & $~~~~1.685~~~$ \\
$A_2$:5 & 0.0772 & 0.0034 & $~~{\bf 22.694}^{***}$ \\
\hline
\end{tabular}
\begin{flushleft}
The intercept terms are omitted. ${}^{*}<.05$, ${}^{**}<.01$, and ${}^{***}<.001$ 
\end{flushleft}
\label{tab:tk_case1}
\end{table}

\begin{table}[!ht]
\centering
\caption{{\bf Species identification in the TK model using logistic regression (Case 2a).} \small Symbol ``+'' indicates that estimated coefficients are significant and have positive signs. Symbols 
\checkmark and \xmark\, indicate correct and incorrect identification of the corresponding reaction, respectively.}
\begin{tabular}{|c|l|lll|r|}
\hline
No. & \multicolumn{1}{c|}{Reactions} & $A_1$ & $A_2$ &  & \# of Obs. \\\hline\hline
1 & $A_1+A_2 \to 2A_1$             & + & + & \checkmark & 367,218\\
2 & $A_1+A_2 \to 2A_2$             & + & + & \checkmark & 386,551 \\
3 & $A_1 \to \emptyset$            & + &   & \checkmark & 1,016\\
4 & $\emptyset \to A_1$            & + &   & \xmark     & 19,918\\
5 & $A_2 \to \emptyset$            &   & + & \checkmark & 28,441\\
6 & $\emptyset \to A_2$ (reference)&   &   &            & 10,022\\\hline\hline
\end{tabular}
\label{table:tk_case2a}
\vspace{.25in}
\caption{{\bf Multinomial logistic regression model fitting summary table for the TK model with asymmetric reaction rate (Case 2a).} \small \ Highly significant positive values are shown in bold. Refer also to the histogram of the 
$z$-values in Fig~\ref{fig:histo_tk2a}.}
\begin{tabular}{crrr}
\hline Reactant:Reaction
 & Estimate & Std. Error & $Z$-value \\
\hline
$A_1$:1 & 0.7049 & 0.0070 & ${\bf 100.118}^{***}$ \\
$A_1$:2 & 0.7047 & 0.0070 & ${\bf 100.100}^{***}$ \\
$A_1$:3 & 0.7208 & 0.0074 & $~~{\bf 97.696}^{***}$ \\
$A_1$:4 & 0.0323 & 0.0083 & $~~~~{\bf 3.915}^{***}$ \\
$A_1$:5 & -3.9053 & 0.0484 & $-80.753^{***}$ \\
$A_2$:1 & 0.0139 & 0.0008 & $~~{\bf 18.174}^{***}$ \\
$A_2$:2 & 0.0137 & 0.0008 & $~~{\bf 17.934}^{***}$ \\
$A_2$:3 & -0.4380 & 0.0147 & $-29.825^{***}$ \\
$A_2$:4 & 0.0007 & 0.0009 & $0.855~~~$ \\
$A_2$:5 & 0.0188 & 0.0009 & $~~{\bf 20.219}^{***}$ \\
\hline
\end{tabular}
\begin{flushleft}
The intercept terms are omitted. ${}^{*}<.05$, ${}^{**}<.01$, and ${}^{***}<.001$ 
\end{flushleft}
\label{tab:tk_case2a}
\end{table}
\begin{table}[!ht]
\centering
\caption{{\bf Species identification in the TK model using logistic regression (Case 2b) with 20 trajectories.} \small  \   Symbol ``+'' indicates that estimated coefficients are significant and have positive signs. 
Symbol \checkmark\ indicates correct identification of the corresponding reaction.}
\begin{tabular}{|c|l|lll|r|}
\hline
No. & \multicolumn{1}{c|}{Reactions} & $A_1$ & $A_2$ &  & \# of Obs. \\\hline\hline
1 & $A_1+A_2 \to 2A_1$             & + & + & \checkmark & 740,653\\
2 & $A_1+A_2 \to 2A_2$             & + & + & \checkmark & 779,731 \\
3 & $A_1 \to \emptyset$            & + &   & \checkmark & 1,817\\
4 & $\emptyset \to A_1$            &   &   & \checkmark & 39,974\\
5 & $A_2 \to \emptyset$            &   & + & \checkmark & 57,317\\
6 & $\emptyset \to A_2$ (reference)&   &   &            & 20,109\\\hline\hline
\end{tabular}
\label{table:tk_case2b}%
\vspace{.25in}
\caption{{\bf Multinomial logistic regression model fitting summary table for the TK model with asymmetric reaction rate using 20 trajectories.}\small \ Highly significant positive values are shown in bold. Refer also to the histogram of $z$-values in Fig~\ref{fig:histo_tk2b}.}
\begin{tabular}{crrr}
\hline
Reactant:Reaction & Estimate & Std. Error & $Z$-value \\
\hline
$A_1$:1     &  0.7789 & 0.0545 & ${\bf 142.813}^{***}$ \\
$A_1$:2     &  0.7786 & 0.0545 & ${\bf 142.772}^{***}$ \\
$A_1$:3     &  0.7908 & 0.0564 & ${\bf 140.258}^{***}$ \\
$A_1$:4     & -0.0042 & 0.0065 & $-0.651$~~~ \\
$A_1$:5     & -3.7930 & 0.0329 & $-115.419^{***}$ \\
$A_2$:1     &  0.0132 & 0.0006 & $~~{\bf 23.480}^{***}$ \\
$A_2$:2     &  0.0130 & 0.0006 & $~~{\bf 23.189}^{***}$ \\
$A_2$:3     & -0.0424 & 0.0101 & $-42.178^{***}$ \\
$A_2$:4     &  0.0442 & 0.0064 & $0.069$~~~ \\
$A_2$:5     &  0.0184 & 0.0068 & $~~{\bf 27.123}^{***}$ \\
\hline
\end{tabular}
\begin{flushleft}
The intercept terms are omitted. ${}^{*}<.05$, ${}^{**}<.01$, and ${}^{***}<.001$ 
\end{flushleft}
\label{tab:tk_case2b}
\end{table}

\begin{table}[!ht]
\centering
\caption{
{\bf Species identification in the Heat Shock Response (HSR) model using logistic regression (Case 1).}  \small Symbol ``+'' indicates that estimated coefficients are significant and have positive signs. Symbol 
\checkmark\ indicates correct identification of the corresponding reaction.}
\begin{tabular}{|c|l|lllcr|}
\hline
No. & \multicolumn{1}{|l|}{ Reactions} & 
$P_1$ & $P_2$ & $R_1$ &  & \# of Obs. \\
\hline\hline 
1  & $P_1 \to R_1$                        & + &   &   & \checkmark & 18,909 \\
2  & $P_2\to R_1$                         &   & + &   & \checkmark & 27,364 \\
3  & $R_1\to P_2$                         &   &   & + & \checkmark & 35,996 \\
4  & $R_1\to 2R_1$                        &   &   & + & \checkmark & 36,322 \\
5  & $R_1+P_2\to \emptyset$              &   & + & + & \checkmark & 10,068 \\
6  & $R_1 \to \emptyset$                 &   &   & + & \checkmark & 36,679 \\
7  & $P_1\to \emptyset$                  & + &   &   & \checkmark & 19,069 \\
8  & $P_2\to \emptyset$                  &   & + &   & \checkmark & 27,536 \\
9  & $\emptyset\to P_1,\,P_2\to P_1+P_2$ &   & + &   & \checkmark & 37,676 \\
10 & $\emptyset\to P_2,\, P_1\to P_1+P_2$ (reference) & & & & & 28,758 \\
\hline\hline
\end{tabular}
\label{table:heat_shock_case1}
\vspace{.25in}
\caption{{\bf Multinomial logistic regression model fitting summary table for the Heat Shock model for Case 1.}\small \ Highly significant positive values are shown in bold. Refer also to the histogram of $z$-values in Fig~\ref{fig:histo_hs1}.}
\begin{tabular}{crrr}
\hline Reactant:Reaction
 & Estimate & Std. Error & $Z$-value \\
\hline
$P_1$:1 & $1.530 \times 10^{-2}$ & $1.888 \times 10^{-3}$ & ${\bf 8.104}^{***}$ \\
$P_1$:2 & $-3.390 \times 10^{-2}$ & $1.747 \times 10^{-3}$ & $-19.402^{***}$ \\
$P_1$:3 & $-3.263 \times 10^{-2}$ & $1.647 \times 10^{-3}$ & $-19.806^{***}$ \\
$P_1$:4 & $-3.236 \times 10^{-2}$ & $1.646 \times 10^{-3}$ & $-19.663^{***}$ \\
$P_1$:5 & $-3.491 \times 10^{-2}$ & $2.351 \times 10^{-3}$ & $-14.852^{***}$ \\
$P_1$:6 & $-3.316 \times 10^{-2}$ & $1.643 \times 10^{-3}$ & $-20.186^{***}$ \\
$P_1$:7 & $1.185 \times 10^{-2}$ & $1.886 \times 10^{-3}$ & ${\bf 6.283}^{***}$ \\
$P_1$:8 & $-3.553 \times 10^{-2}$ & $1.746 \times 10^{-3}$ & $-20.353^{***}$ \\
$P_1$:9 & $-3.245 \times 10^{-2}$ & $1.625 \times 10^{-3}$ & $-19.971^{***}$ \\
$P_2$:1 & $-3.430 \times 10^{-3}$ & $1.735 \times 10^{-3}$ & $-1.977^{*}~~$ \\
$P_2$:2 & $3.368 \times 10^{-2}$ & $1.554 \times 10^{-3}$ & ${\bf 21.677}^{***}$ \\
$P_2$:3 & $-6.268 \times 10^{-5}$ & $1.469 \times 10^{-3}$ & $-0.043$~~~ \\
$P_2$:4 & $-1.564 \times 10^{-3}$ & $1.467 \times 10^{-3}$ & $-1.066$~~~ \\
$P_2$:5 & $3.432 \times 10^{-2}$ & $2.107 \times 10^{-3}$ & ${\bf 16.285}^{***}$ \\
$P_2$:6 & $-1.053 \times 10^{-3}$ & $1.464 \times 10^{-3}$ & $-0.719$~~~ \\
$P_2$:7 & $-3.781 \times 10^{-4}$ & $1.729 \times 10^{-3}$ & $-0.219$~~~ \\
$P_2$:8 & $3.458 \times 10^{-2}$ & $1.551 \times 10^{-3}$ & ${\bf 22.297}^{***}$ \\
$P_2$:9 & $2.479 \times 10^{-2}$ & $1.446 \times 10^{-3}$ & ${\bf 17.143}^{***}$ \\
$R_1$:1 & $1.924 \times 10^{-3}$ & $1.070 \times 10^{-3}$ & $1.798~~~$ \\
$R_1$:2 & $1.268 \times 10^{-3}$ & $9.705 \times 10^{-4}$ & $1.307$ ~~~\\
$R_1$:3 & $2.850 \times 10^{-2}$ & $8.913 \times 10^{-4}$ & ${\bf 31.975}^{***}$ \\
$R_1$:4 & $2.774 \times 10^{-2}$ & $8.901 \times 10^{-4}$ & ${\bf 31.162}^{***}$ \\
$R_1$:5 & $2.671 \times 10^{-2}$ & $1.283 \times 10^{-3}$ & ${\bf 20.817}^{***}$ \\
$R_1$:6 & $2.747 \times 10^{-2}$ & $8.885 \times 10^{-4}$ & ${\bf 30.919}^{***}$ \\
$R_1$:7 & $1.601 \times 10^{-3}$ & $1.068 \times 10^{-3}$ & $1.499$~~~ \\
$R_1$:8 & $1.838 \times 10^{-3}$ & $9.686 \times 10^{-4}$ & $1.898~~~$ \\
$R_1$:9 & $1.553 \times 10^{-3}$ & $8.997 \times 10^{-4}$ & $1.726$~~~ \\
\hline
\end{tabular}
\begin{flushleft}
The intercept terms are omitted. ${}^{*}<.05$, ${}^{**}<.01$, and ${}^{***}<.001$ 
\end{flushleft}
\label{tab:hs_case1}
\end{table}

\begin{table}[!ht]
\centering
\caption{
{\bf Species identification in the Heat Shock  model using logistic regression (Case 2a).}\small \  Symbol ``+'' indicates that estimated coefficients are significant and have positive signs. Symbol
\checkmark\ indicates correct identification of the corresponding reaction.}
\begin{tabular}{|c|l|lllc|r|}
\hline
No. & \multicolumn{1}{|l|}{ Reactions} & 
$P_1$ & $P_2$ & $R_1$ &  & \# of Obs. \\
\hline\hline 
1  & $P_1 \to R_1$                        &   &   &   &          & 14,793 \\
2  & $P_2\to R_1$                         &   & + &   & \checkmark & 18,953 \\
3  & $R_1\to P_2$                         &   &   & + & \checkmark & 28,571 \\
4  & $R_1\to 2R_1$                        &   &   & + & \checkmark & 28,763 \\
5  & $R_1+P_2\to \emptyset$              &   & + & + & \checkmark & 5,749  \\
6  & $R_1 \to \emptyset$                 &   &   & + & \checkmark & 28,411 \\
7  & $P_1\to \emptyset$                  &   &   &   &            & 14,847 \\
8  & $P_2\to \emptyset$                  &   & + &   & \checkmark     & 18,997 \\
9  & $\emptyset\to P_1,\,P_2\to P_1+P_2$ &   & + &   & \checkmark     & 29,252 \\
10 & $\emptyset\to P_2,\, P_1\to P_1+P_2$ (reference) & & & & & 14,805 \\
\hline\hline
\end{tabular}
\label{table:heat_shock_case2a}
\vspace{.25in}
\caption{{\bf Multinomial logistic regression model fitting summary table for the Heat Shock model for Case 2.}\small \ Highly significant positive values are shown in bold. Refer also to the histogram of $z$-values in Fig~\ref{fig:histo_hs2a}.}
\begin{tabular}{crrr}
\hline Reactant:Reaction
 & Estimate & Std. Error & $Z$-value \\
\hline
$P_1$:1 & $-3.141 \times 10^{-3}$ & $2.567 \times 10^{-3}$ & $-1.224~~~$ \\
$P_1$:2 & $-6.356 \times 10^{-2}$ & $2.479 \times 10^{-3}$ & $-25.638^{***}$ \\
$P_1$:3 & $-6.109 \times 10^{-2}$ & $2.304 \times 10^{-3}$ & $-26.515^{***}$ \\
$P_1$:4 & $-6.130 \times 10^{-2}$ & $2.301 \times 10^{-3}$ & $-26.645^{***}$ \\
$P_1$:5 & $-6.237 \times 10^{-2}$ & $3.388 \times 10^{-3}$ & $-18.409^{***}$ \\
$P_1$:6 & $-5.844 \times 10^{-2}$ & $2.302 \times 10^{-3}$ & $-25.390^{***}$ \\
$P_1$:7 & $-1.075 \times 10^{-3}$ & $2.565 \times 10^{-3}$ & $-0.419~~~$ \\
$P_1$:8 & $-5.759 \times 10^{-2}$ & $2.471 \times 10^{-3}$ & $-23.307^{***}$ \\
$P_1$:9 & $-5.932 \times 10^{-2}$ & $2.292 \times 10^{-3}$ & $-25.882^{***}$ \\
$P_2$:1 & $1.112 \times 10^{-3}$ & $2.473 \times 10^{-3}$ & $0.450~~~$ \\
$P_2$:2 & $5.160 \times 10^{-2}$ & $2.321 \times 10^{-3}$ & ${\bf 22.231}^{***}$ \\
$P_2$:3 & $5.130 \times 10^{-4}$ & $2.167 \times 10^{-3}$ & $0.237~~~$ \\
$P_2$:4 & $1.928 \times 10^{-3}$ & $2.136 \times 10^{-3}$ & $0.891~~~$ \\
$P_2$:5 & $4.977 \times 10^{-2}$ & $3.213 \times 10^{-3}$ & ${\bf 15.492}^{***}$ \\
$P_2$:6 & $1.170 \times 10^{-5}$ & $2.168 \times 10^{-3}$ & $0.098~~~$ \\
$P_2$:7 & $3.188 \times 10^{-4}$ & $2.471 \times 10^{-3}$ & $0.129~~~$ \\
$P_2$:8 & $4.641 \times 10^{-2}$ & $2.320 \times 10^{-3}$ & ${\bf 20.002}^{***}$ \\
$P_2$:9 & $3.155 \times 10^{-2}$ & $2.149 \times 10^{-3}$ & ${\bf 14.679}^{***}$ \\
$R_1$:1 & $3.888 \times 10^{-3}$ & $1.441 \times 10^{-3}$ & $2.698^{**}~$ \\
$R_1$:2 & $2.490 \times 10^{-3}$ & $1.364 \times 10^{-3}$ & $1.826~~~$ \\
$R_1$:3 & $3.673 \times 10^{-2}$ & $1.251 \times 10^{-3}$ & ${\bf 29.367}^{***}$ \\
$R_1$:4 & $3.565 \times 10^{-2}$ & $1.250 \times 10^{-3}$ & ${\bf 28.523}^{***}$ \\
$R_1$:5 & $3.339 \times 10^{-2}$ & $1.865 \times 10^{-3}$ & ${\bf 17.903}^{***}$ \\
$R_1$:6 & $3.642 \times 10^{-2}$ & $1.252 \times 10^{-3}$ & ${\bf 29.099}^{***}$ \\
$R_1$:7 & $2.164 \times 10^{-3}$ & $1.441 \times 10^{-3}$ & $1.502~~~$ \\
$R_1$:8 & $3.247 \times 10^{-3}$ & $1.362 \times 10^{-3}$ & $2.384^{*}~~$ \\
$R_1$:9 & $3.587 \times 10^{-3}$ & $1.257 \times 10^{-3}$ & $2.853^{**}~$ \\
\hline
\end{tabular}
\begin{flushleft}
The intercept terms are omitted. ${}^{*}<.05$, ${}^{**}<.01$, and ${}^{***}<.001$
\end{flushleft}
\label{tab:hs_case2a}
\end{table}

\begin{table}[!ht]
\centering
\caption{
{\bf Species identification in the Heat Shock  model using logistic regression (Case 2b) with 20 trajectories.} \small Symbol ``+'' indicates that estimated coefficients are significant and have positive signs. Symbol 
\checkmark\ indicates correct identification of the corresponding reaction.}
\begin{tabular}{|c|l|lllc|r|}
\hline
No. & \multicolumn{1}{|l|}{ Reactions} & 
$P_1$ & $P_2$ & $R_1$ &  & \# of Obs. \\
\hline\hline 
1  & $P_1 \to R_1$                        &   &   &  &      & 29,418 \\
2  & $P_2\to R_1$                         &   & + &   & \checkmark & 37,678 \\
3  & $R_1\to P_2$                         &   &   & + & \checkmark & 56,627 \\
4  & $R_1\to 2R_1$                        &   &  & + & \checkmark & 57,162 \\
5  & $R_1+P_2\to \emptyset$              &   & + & + & \checkmark & 11,310  \\
6  & $R_1 \to \emptyset$                 &   &   & + & \checkmark & 56,781 \\
7  & $P_1\to \emptyset$                  &   &   &   &            & 29,410 \\
8  & $P_2\to \emptyset$                  &   & + & & \checkmark     & 37,800 \\
9  & $\emptyset\to P_1,\,P_2\to P_1+P_2$ &   & + & & \checkmark     & 58,089 \\
10 & $\emptyset\to P_2,\, P_1\to P_1+P_2$ (reference) & & & & & 29,531 \\
\hline\hline
\end{tabular}
\label{table:heat_shock_case2b}
\vspace{.25in}
\caption{{\bf Multinomial logistic regression model fitting summary table for the Heat Shock model for Case 2 using 20 trajectories.}\small \ Highly significant positive values are shown in bold. Refer also to the histogram of $z$-values in Fig~\ref{fig:histo_hs2b}.}
\begin{tabular}{crrr}
\hline
Reactant:Reaction & Estimate & Std. Error & $Z$-value \\
\hline
$P_{1}$:1 & -0.015178 & 0.0018450 & $-0.823$~~~ \\
$P_{1}$:2 & -0.0635461 & 0.0017824 & $-35.653^{***}$ \\
$P_{1}$:3 & -0.0612178 & 0.0016570 & $-36.946^{***}$ \\
$P_{1}$:4 & -0.0608066 & 0.0016527 & $-36.792^{***}$ \\
$P_{1}$:5 & -0.0632578 & 0.0024558 & $-25.759^{***}$ \\
$P_{1}$:6 & -0.0590388 & 0.0016548 & $-35.677^{***}$ \\
$P_{1}$:7 & -0.011454 & 0.0018422 & $-0.622$~~~ \\
$P_{1}$:8 & -0.0588260 & 0.0017747 & $-33.147^{***}$ \\
$P_{1}$:9 & -0.0601327 & 0.0016462 & $-36.529^{***}$ \\
$P_{2}$:1 & 0.0009753 & 0.0017539 & $0.556$~~~ \\
$P_{2}$:2 & 0.0513635 & 0.0016457 & ${\bf 31.210}^{***}$ \\
$P_{2}$:3 & 0.028941 & 0.0015372 & $1.883~~~$ \\
$P_{2}$:4 & 0.031959 & 0.0015342 & $2.083^{*}~~$ \\
$P_{2}$:5 & 0.0504529 & 0.0022867 & ${\bf 22.064}^{***}$ \\
$P_{2}$:6 & -0.0001088 & 0.0015370 & $-0.071$~~~ \\
$P_{2}$:7 & 0.026028 & 0.0017525 & $1.485$~~~ \\
$P_{2}$:8 & 0.0495787 & 0.0016438 & ${\bf 30.161}^{***}$ \\
$P_{2}$:9 & 0.0339964 & 0.0015239 & ${\bf 22.308}^{***}$ \\
$R_{1}$:1 & 0.0007089 & 0.0010109 & $0.701$~~~ \\
$R_{1}$:2 & 0.0009651 & 0.0009560 & $1.009$~~~ \\
$R_{1}$:3 & 0.0335541 & 0.0008757 & ${\bf 38.316}^{***}$ \\
$R_{1}$:4 & 0.0341119 & 0.0008741 & ${\bf 39.026}^{***}$ \\
$R_{1}$:5 & 0.0326457 & 0.0013078 & ${\bf 24.962}^{***}$ \\
$R_{1}$:6 & 0.0351653 & 0.0008749 & ${\bf 40.193}^{***}$ \\
$R_{1}$:7 & 0.0002578 & 0.0010109 & $0.255$~~~ \\
$R_{1}$:8 & 0.0008692 & 0.0009548 & $0.910$~~~ \\
$R_{1}$:9 & 0.0013080 & 0.0008812 & $1.484$~~~ \\
\hline
\end{tabular}
\begin{flushleft}
The intercept terms are omitted.  ${}^{*}<.05$, ${}^{**}<.01$, and ${}^{***}<.001$
\end{flushleft}
\label{tab:hs_case2b}
\end{table}

\begin{table}[!ht]
\centering
\caption{
{\bf Species identification in the SIR model with demography using logistic regression.} \small Symbol ``+'' indicates that estimated coefficients are significant and have positive signs. Symbol 
\checkmark\ indicates correct identification of the corresponding reaction.}
\begin{tabular}{|c|l|lllc|r|}
\hline
No. & \multicolumn{1}{|l|}{Reactions} & $S$ & $I$ & $R$ &  & \# of Obs. \\
\hline\hline
1 & $S+I \to 2I$             & + & + &   & \checkmark & 3,248 \\
2 & $I \to R$                &   & + &   & \checkmark & 3,105 \\
3 & $S \to \emptyset$        & + &   &   & \checkmark & 60   \\
4 & $I \to \emptyset$        &   & + &   & \checkmark & 15   \\
5 & $R \to \emptyset$        &   &   & + & \checkmark & 174  \\
6 & $\emptyset \to S$ (reference) & & & & & 3,033 \\
\hline\hline
\end{tabular}

\label{table:sir2}
\vspace{.25in}
\caption{{\bf Multinomial logistic regression model fitting summary table for the SIR model.} \small \ Highly significant positive values are shown in bold. Refer also to the histogram of the $z$-values in Fig~\ref{fig:histo_sir}.}
\begin{tabular}{crrr}
\hline Reactant:Reaction
 & Estimate & Std. Error & $z$-value \\
\hline
$S$:1 & $4.669 \times 10^{-3}$ & $1.117 \times 10^{-3}$ & ${\bf 4.179}^{***}$ \\
$S$:2 & $-1.113 \times 10^{-2}$ & $1.259 \times 10^{-3}$ & $-8.837^{***}$ \\
$S$:3 & $1.490 \times 10^{-2}$ & $4.108 \times 10^{-3}$ & ${\bf 3.626}^{***}$ \\
$S$:4 & $-7.696 \times 10^{-3}$ & $1.342 \times 10^{-2}$ & $-0.574~~~$ \\
$S$:5 & $-6.200 \times 10^{-3}$ & $3.638 \times 10^{-3}$ & $-1.704~~~$ \\
$I$:1 & $5.822 \times 10^{-2}$ & $2.486 \times 10^{-3}$ & ${\bf 23.417}^{***}$ \\
$I$:2 & $5.878 \times 10^{-2}$ & $2.546 \times 10^{-3}$ & ${\bf 23.087}^{***}$ \\
$I$:3 & $-3.624 \times 10^{-2}$ & $1.704 \times 10^{-2}$ & $-2.127^{*}~~~$ \\
$I$:4 & $8.910 \times 10^{-2}$ & $2.220 \times 10^{-2}$ & ${\bf 4.013}^{***}$ \\
$I$:5 & $9.489 \times 10^{-4}$ & $8.675 \times 10^{-3}$ & $0.109~~~$ \\
$R$:1 & $-2.883 \times 10^{-5}$ & $3.376 \times 10^{-4}$ & $-0.085~~~$ \\
$R$:2 & $4.775 \times 10^{-5}$ & $3.566 \times 10^{-4}$ & $0.134~~~$ \\
$R$:3 & $-1.922 \times 10^{-3}$ & $1.460 \times 10^{-3}$ & $-1.317~~~$ \\
$R$:4 & $-5.789 \times 10^{-4}$ & $3.949 \times 10^{-3}$ & $-0.147~~~$ \\
$R$:5 & $5.000 \times 10^{-3}$ & $1.043 \times 10^{-3}$ & ${\bf 4.793}^{***}$ \\
\hline
\end{tabular}
\begin{flushleft}
The intercept terms are omitted. ${}^{*}<.05$, ${}^{**}<.01$, and ${}^{***}<.001$ 
\end{flushleft}
\label{tab:sir3}
\end{table}

\clearpage
\section*{Figures}
\ 
\begin{figure}[h!]
\includegraphics[width=1.0\textwidth]{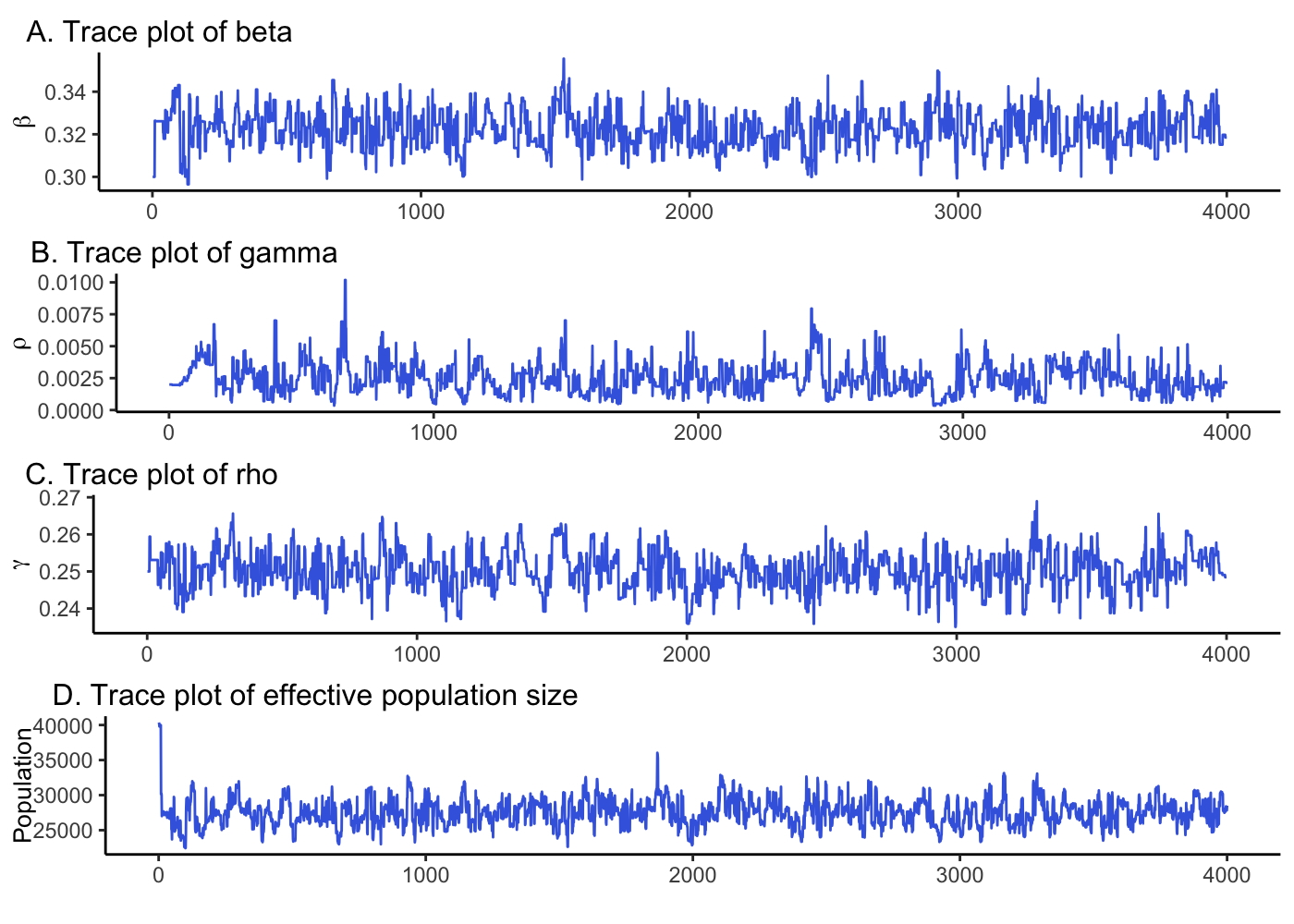}
\caption{Trace plots of the posterior sample of the parameters $\theta =(\beta, \gamma, \rho)$ and effective population size $n$.}
\label{fig:trace}
\end{figure}

\begin{figure}
\includegraphics[width=1.0\textwidth]{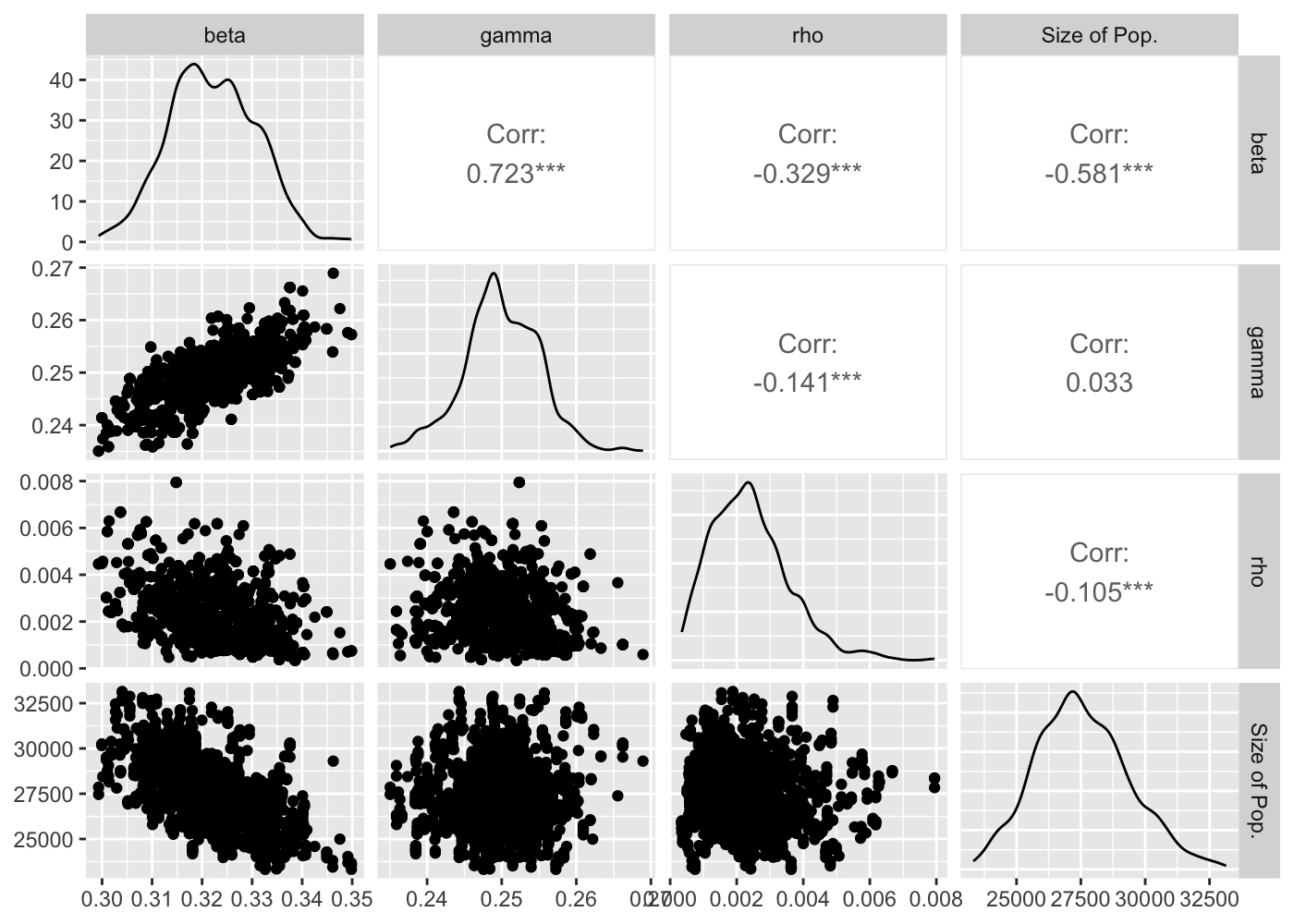}
\caption{Posterior paired scatter plots and histograms of the parameters. The off-diagonal terms represent the scatter plots between each parameters and the diagonal terms are the histogram of the parameters.}
\label{fig:corr}
\end{figure}

\begin{figure}
\includegraphics[width=1.0\textwidth]{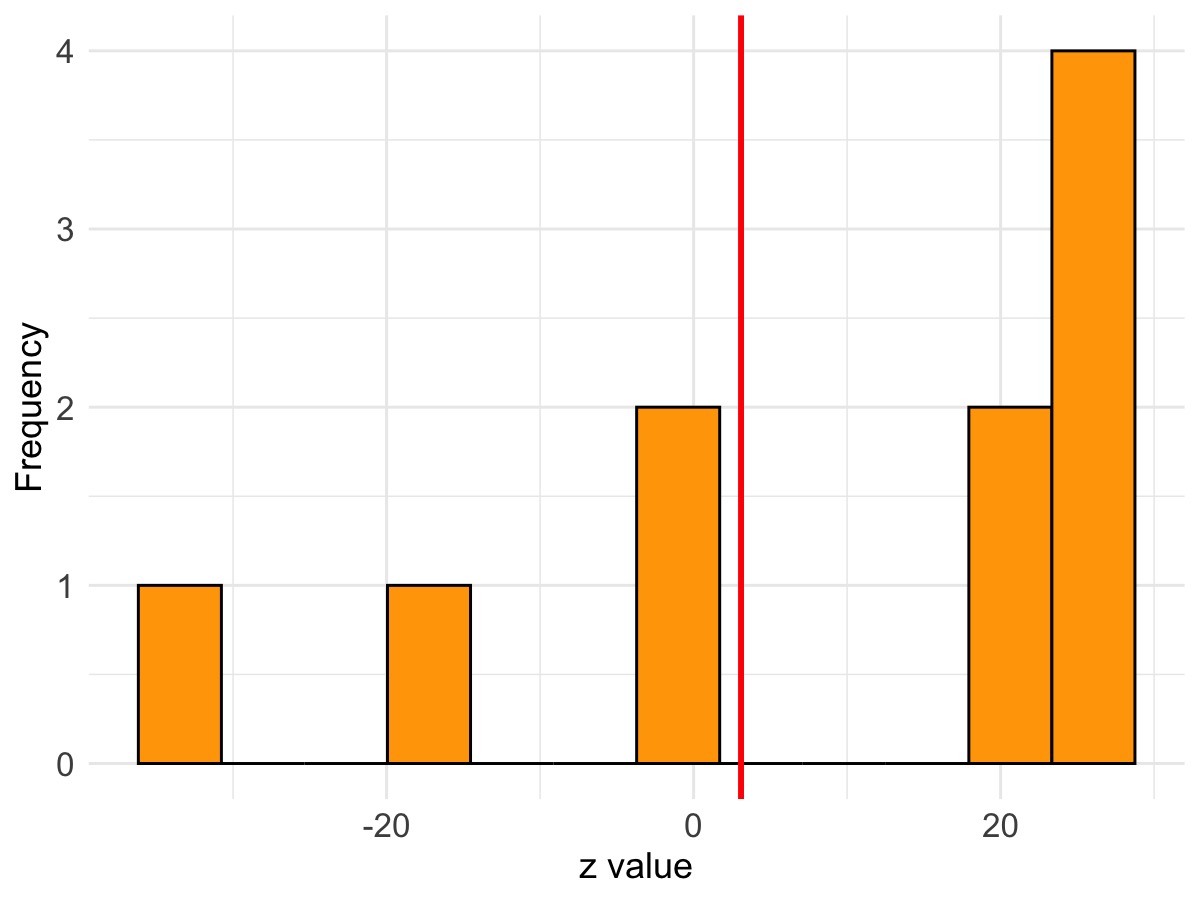}
\caption{\textbf{Histogram of $z$-values for the TK model under symmetric reaction rates (Case 1)}. The red vertical line denotes the threshold $z$-value of 3.09, corresponding to the criterion for coefficient significance ($P<0.001$).}
\label{fig:histo_tk1}
\end{figure} 

\begin{figure}
\includegraphics[width=1.0\textwidth]{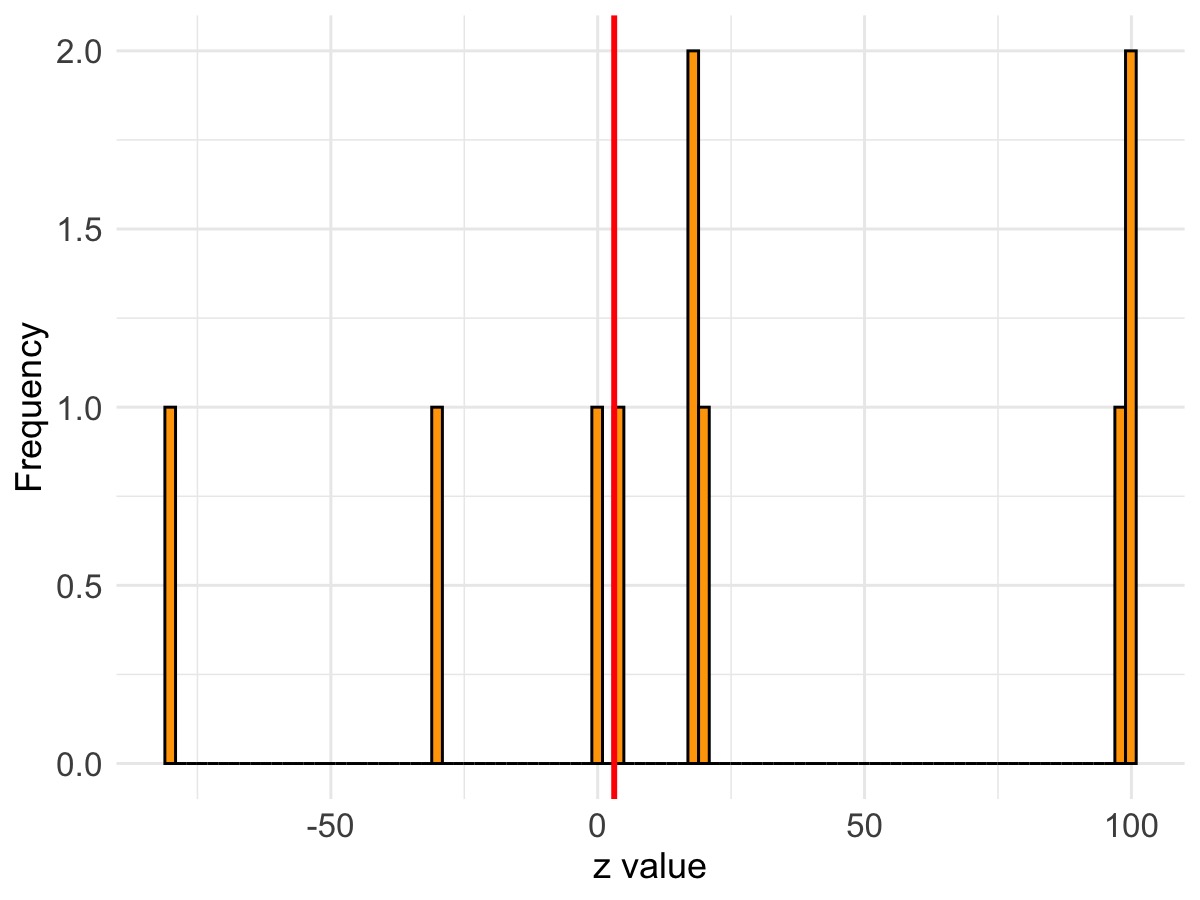}
\caption{\textbf{Histogram of the $z$-values for the TK model asymmetric reaction rate (Case 2a).} The red vertical line denotes the threshold $z$-value of 3.09, corresponding to the criterion for coefficient significance ($P<0.001$).}\label{fig:histo_tk2a}
\end{figure} 

\begin{figure}
\includegraphics[width=1.0\textwidth]{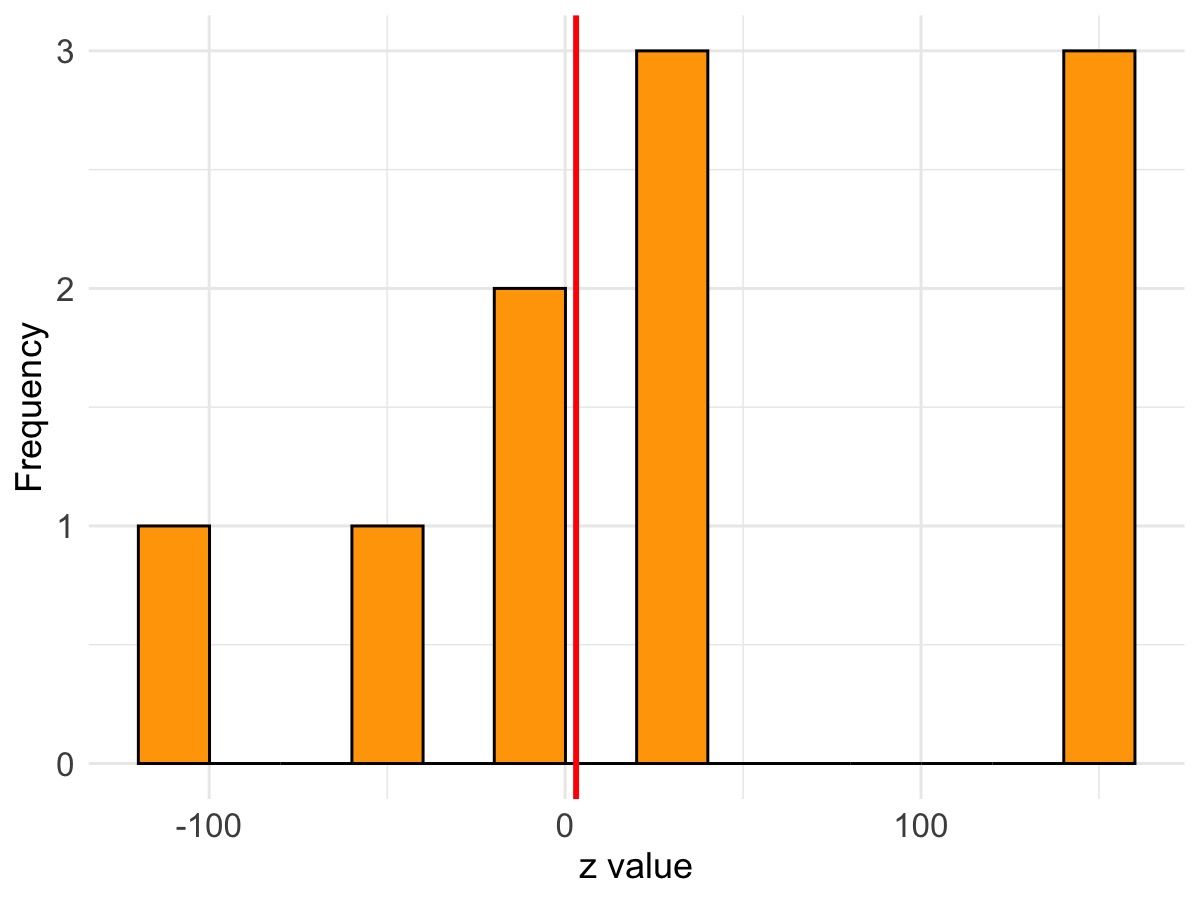}
\caption{\textbf{Histogram of the $z$-values for the TK model asymmetric reaction rate using 20 trajectories (Case 2b).} The red vertical line denotes the threshold $z$-value of 3.09, corresponding to the criterion for coefficient significance ($P<0.001$).}
\label{fig:histo_tk2b}
\end{figure} 

\begin{figure}
\includegraphics[width=1.0\textwidth]{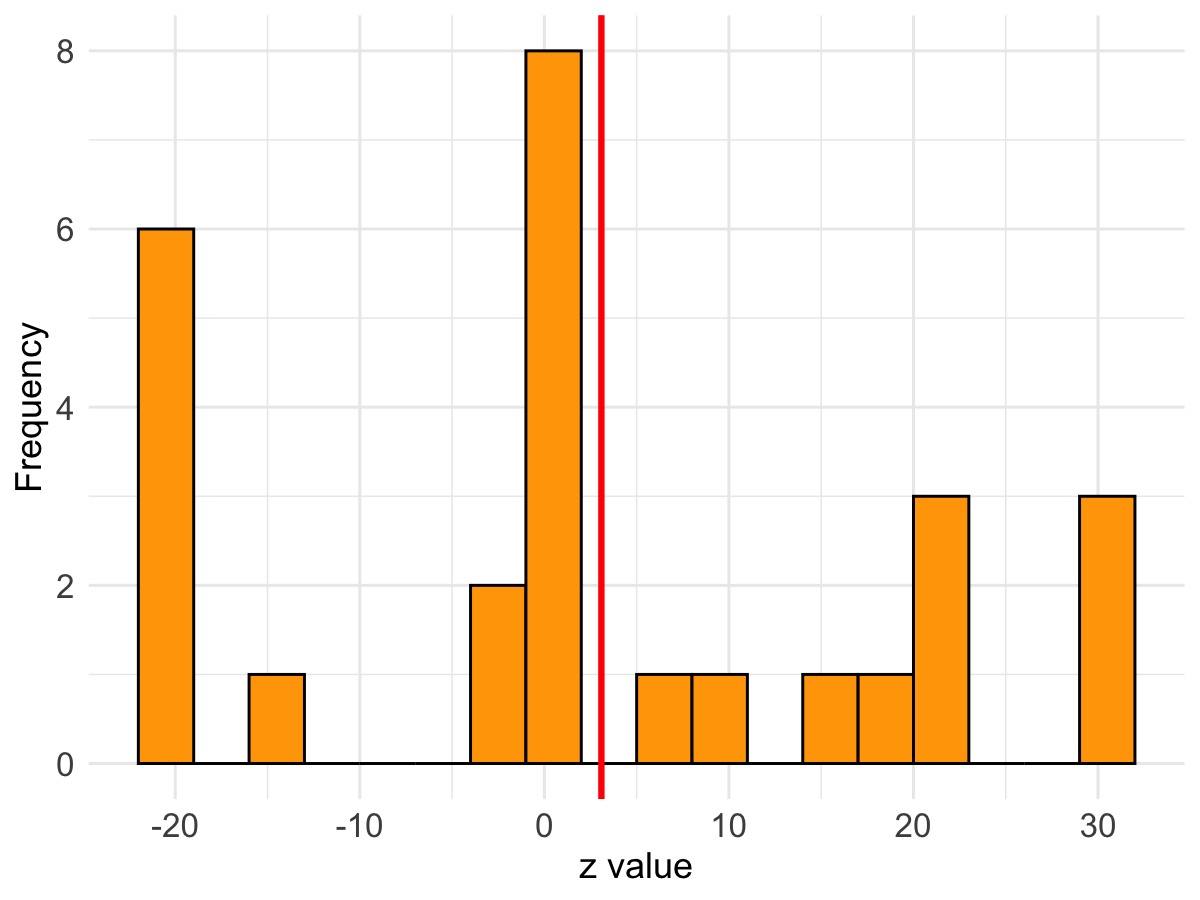}
\caption{\textbf{Histogram of the $z$-values for the Heat Shock Response model for Case 1.} The red vertical line denotes the threshold $z$-value of 3.09, corresponding to the criterion for coefficient significance ($P<0.001$).}
\label{fig:histo_hs1}
\end{figure} 

\begin{figure}
\includegraphics[width=1.0\textwidth]{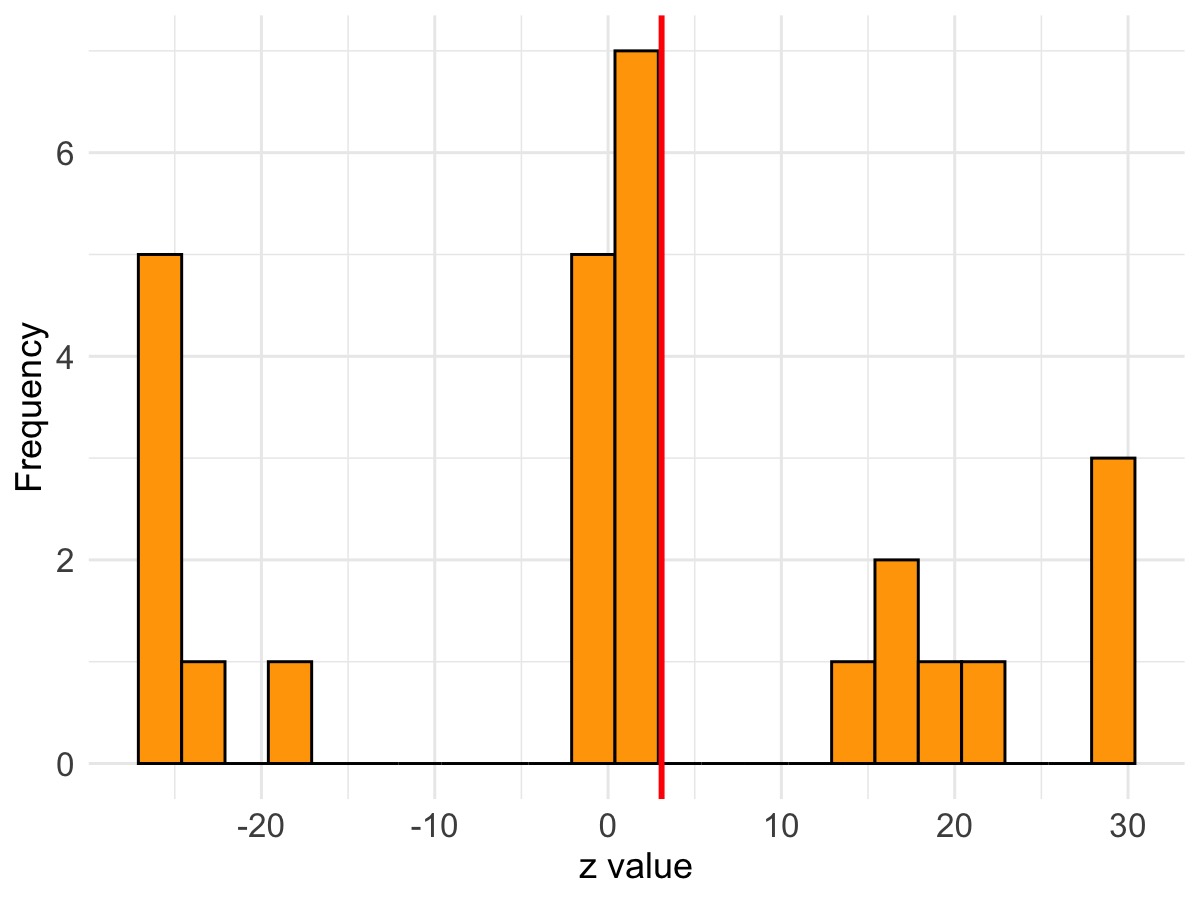}
\caption{\textbf{Histogram of the $z$-values for the Heat Shock model for Case 2a.} The red vertical line denotes the threshold $z$-value of 3.09, corresponding to the criterion for coefficient significance ($P<0.001$).}
\label{fig:histo_hs2a}
\end{figure} 

\begin{figure}
\includegraphics[width=1.0\textwidth]{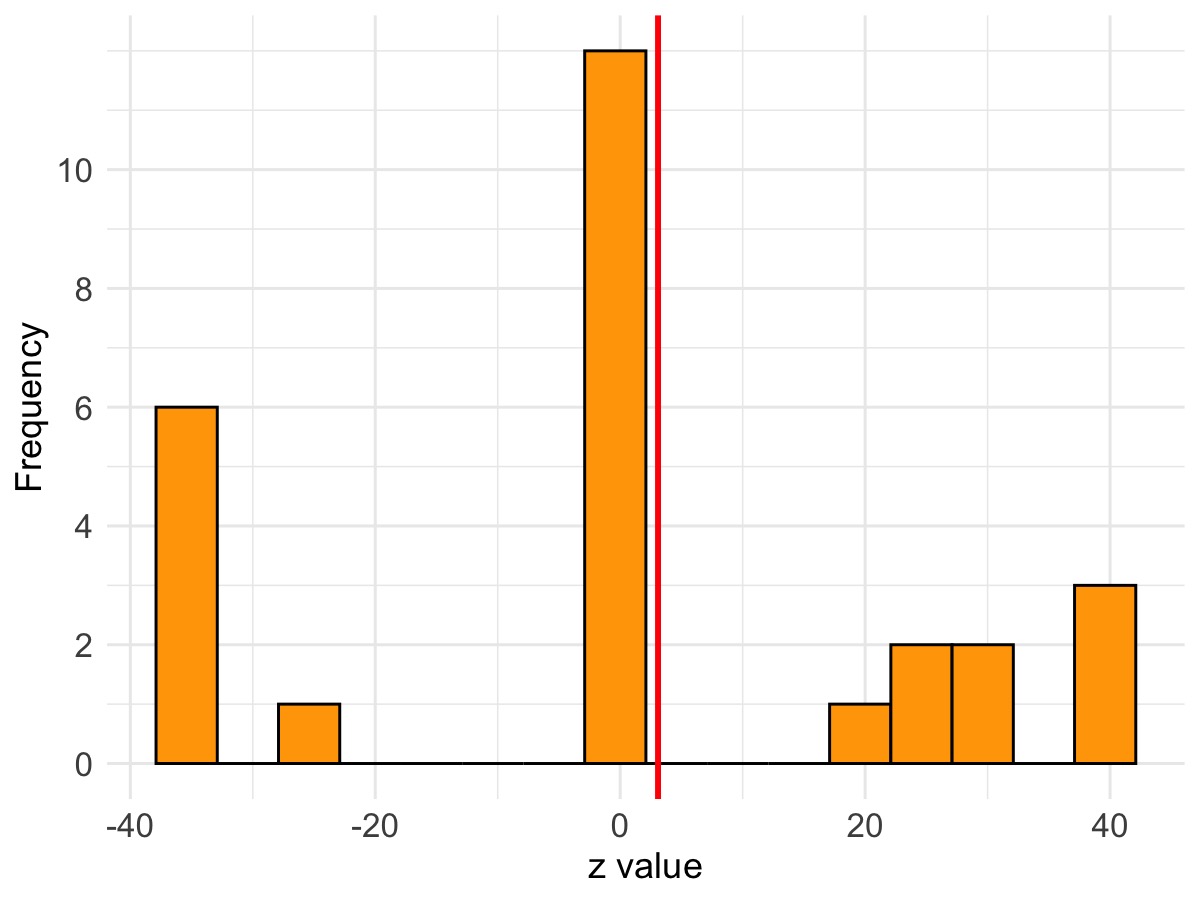}
\caption{\textbf{Histogram of the $z$-values for the Heat Shock model for Case 2b using 20 trajectories.} The red vertical line denotes the threshold $z$-value of 3.09, corresponding to the criterion for coefficient significance ($P<0.001$).}
\label{fig:histo_hs2b}
\end{figure} 

\begin{figure}
\includegraphics[width=1.0\textwidth]{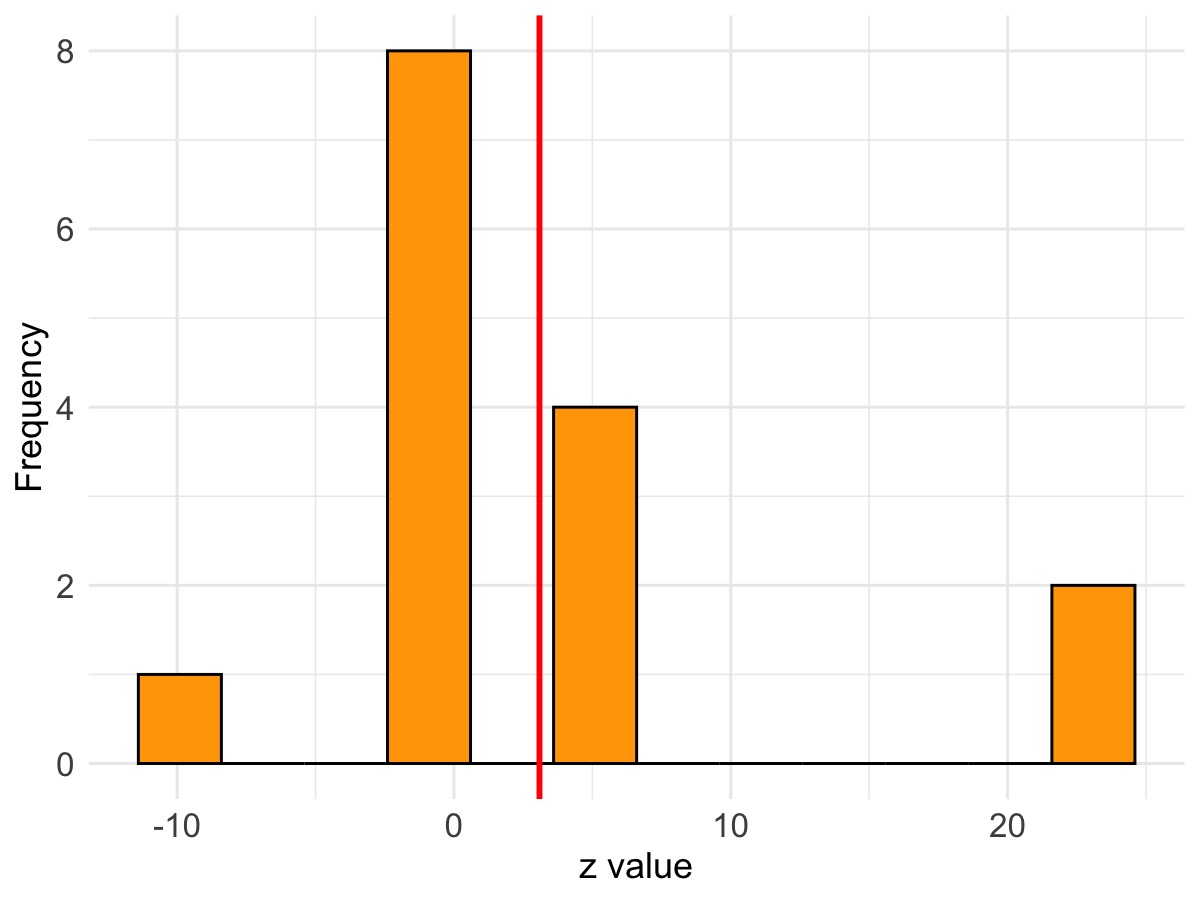}
\caption{\textbf{Histogram for the $z$-values for the logistic regression coefficients in SIR model.} The red vertical line denotes the threshold $z$-value of 3.09, corresponding to the criterion for coefficient significance ($P<0.001$). 
}
\label{fig:histo_sir}
\end{figure} 

\clearpage
\section*{Supplemental Material: Additional Model Details}
 \label{S1_Appendix} 
 \subsection*{Stochastic SIR Model with Demography }

Let $ Y_i(t) $ for $ i = 1, \dots, 6 $ be independent unit-rate Poisson processes. The evolution of the susceptible $ S(t) $, infected $ I(t) $, and recovered $ R(t) $ populations is given by (see, for instance, \cite{Anderson:2015:SAB})

\[
\begin{aligned}
S(t) &= S(0) + Y_1(\mu n_0 t) - Y_2\left( \int_0^t \frac{\beta}{n_0} S(s) I(s) \, \differential{s} \right) - Y_4\left( \int_0^t \nu S(s) \, \differential{s} \right), \\
I(t) &= I(0) + Y_2\left( \int_0^t \frac{\beta}{n_0} S(s) I(s) \, \differential{s} \right) - Y_3\left( \int_0^t \gamma I(s) \, \differential{s} \right) - Y_5\left( \int_0^t \nu I(s) \, \differential{s} \right), \\
R(t) &= R(0) + Y_3\left( \int_0^t \gamma I(s) \, \differential{s} \right) - Y_6\left( \int_0^t \nu R(s) \, \differential{s} \right),
\end{aligned}
\]

\[
S(0) = n_0, \quad I(0) = \rho n_0, \quad R(0) = 0,
\]
where $n_0$ denotes the initial total number of susceptible individuals.
Assume that $n_0$ is large and consider the limit as $n_0$ goes to infinity.
The resulting mean field limiting ODE system is then given by \eqref{eq:demsir}.

\subsection*{Epidemic Size and  Inclusion  Probability}
In the mean field limit  we need to calculate the average count  of $Y_2$,  so 
 the formula  for the  unscaled  epidemic size  at time $t$ is given as:
\[\tilde\tau_t = 1-s_t+\mu t -\int_0^t \nu s_u \differential{u}.\]
Since  $\tilde\tau_t$ is not bounded as a function of  $t$,  we need 
 the demography--corrected  formula for the epidemic size   \begin{equation}\label{eq:tau}
     \tau_t = \frac{1-s_t+\mu t -\int_0^t \nu s_u \differential{u}}{1+\mu t -\int_0^t \nu s_u \differential{u}}, \end{equation}
which  has now the correct interpretation,  under the  assumptions that our $t$ is such that the numerator above is positive (this will be always true, for instance, when $\mu>\nu$).  The formula above  may be also interpreted via the limiting argument as follows.  
The initial amount of $S$ is $n_0$, but by time $t$ there has been some changes to the initial population  not due to infections (flow in and flow out of $S$)  therefore the correct relative count should be 
$$ {\cal T} = \frac{Y_2\left( \int_0^t \frac{\beta}{n_0} S(s) I(s) \, \differential{s} \right)}{S(0) + Y_1(\mu n_0 t)  - Y_4\left( \int_0^t \nu S(s) \, \differential{s} \right)}.$$ Now  $\tau_t$ is simply the approximation to ${\cal T}$   for large $n_0$.

\end{document}